\documentclass[a4paper,onecolumn,11pt,accepted=2023-04-05]{quantumarticle}
\pdfoutput=1
\usepackage[utf8]{inputenc}
\usepackage[english]{babel}
\usepackage[T1]{fontenc}
\usepackage{amsmath}
\usepackage{hyperref}
\usepackage{amsfonts}

\usepackage{amssymb}
\usepackage{graphicx}
\usepackage[graphicx]{realboxes}
\usepackage{float}
\usepackage{multirow}
\usepackage{mathtools}
\usepackage{bbm}
\usepackage{braket}
\usepackage{ytableau,tikz,varwidth}
\usepackage[enableskew]{youngtab}
\usepackage{hyphenat}
\usepackage{verbatim}
\usepackage{subcaption}
\usepackage{hyperref}
\usepackage{soul}
\usepackage{etoolbox}
\usepackage{nameref}
\usepackage{array,multirow,graphicx}
\usetikzlibrary{tikzmark,decorations.pathreplacing,calc}
\usepackage{changepage}
\usepackage{bm}
\usepackage{lipsum}
\usepackage{amsthm}

\usepackage[sorting = none, backend = bibtex, style = nature]{biblatex}
\AtEveryBibitem{\clearfield{url}}
\AtEveryBibitem{\clearfield{eprint}}
\AtEveryBibitem{\clearfield{isbn}}
\AtEveryBibitem{\clearfield{month}}

\usepackage{mathtools}

\def\CC {{\mathbb C}}     
\renewcommand{\S}{\mathcal{S}}
\def\mc {\mathcal}
\def\mk {\mathfrak}

\newcommand{\tr}{\mbox{Tr}}

\def\={\;=\;} \def\+{\,+\,}     
      
      \def\d{\delta} 

\newcommand{\sgn}[1]{\mathrm{sgn}(#1)}

\newtheorem{theorem}{Theorem}
\newtheorem{lemma}[theorem]{Lemma}

\newtheorem{corollary}[theorem]{Corollary}

\newtheorem{fact}{Fact}

\newcommand{\E}[1]{\underset{U \sim #1}{\mathbb{E}}}
\newcommand{\Ec}[2]{\underset{U_{#2} \sim #1}{\mathbb{E}}}

\DeclareRobustCommand{\bbone}{\text{\usefont{U}{bbold}{m}{n}1}}

\graphicspath{{plots/}}

\begin{document}

\title{Matrix concentration inequalities and efficiency of random universal sets of quantum gates}

\author{Piotr Dulian}
\affiliation{Center for Theoretical Physics, Polish Academy of Sciences, Al. Lotnik\'ow 32/46, 02-668 Warsaw, Poland}
\affiliation{Faculty of Physics, University of Warsaw, Pasteura 5, 02-093 Warsaw, Poland}

\author{Adam Sawicki}
\affiliation{Center for Theoretical Physics, Polish Academy of Sciences, Al. Lotnik\'ow 32/46, 02-668 Warsaw, Poland}

\maketitle

\begin{abstract}
For a random set $\mathcal{S} \subset U(d)$ of quantum gates we provide bounds on the probability that $\mathcal{S}$ forms a $\delta$-approximate $t$-design. In particular we have found that for $\mathcal{S}$ drawn from an exact $t$-design the probability that it forms a $\delta$-approximate $t$-design satisfies the inequality $\mathbb{P}\left(\delta \geq x \right)\leq 2D_t \, \frac{e^{-|\mc{S}| x \, \mathrm{arctanh}(x)}}{(1-x^2)^{|\mc{S}|/2}} = O\left( 2D_t \left( \frac{e^{-x^2}}{\sqrt{1-x^2}} \right)^{|\mathcal{S}|} \right)$, where $D_t$ is a sum over dimensions of unique irreducible representations appearing in the decomposition of $U \mapsto U^{\otimes t}\otimes \bar{U}^{\otimes t}$. We use our results to show that to obtain a $\delta$-approximate $t$-design with probability $P$ one needs $O( \delta^{-2}(t\log(d)-\log(1-P)))$ many random gates. We also analyze how $\delta$ concentrates around its expected value $\mathbb{E}\delta$ for random $\mc{S}$. Our results are valid for both symmetric and non-symmetric sets of gates.
\end{abstract}

\section{Introduction and summary of main results}
Practical realisations of quantum computers are constricted by noise and decoherence that affect
large-scale quantum systems.  Although quantum error correction codes can overcome those effects they require usage of thousands of physical qubits to implement a single logical noiseless fault-tolerant qubit \cite{fowler}. This is clearly out of reach for contemporary quantum computers with the number of physical qubits of the order of hundreds \cite{preskill}. Hence, in the near future we are forced to work with noisy intermediate-scale quantum devices (NISQ) \cite{preskill, boxio, montanare}. Moreover, currently the best error rates per gate are the order of \textcolor{black}{$0.1\%$} \cite{ballance, barends} which implies we cannot build circuits much longer than thousand \cite{preskill}. The  length of a circuit is also limited by the coherence time and the time of execution of a single gate. It is noteworthy that it is hard to make gates that are both fast and have low error rates \cite{ballance}. Clearly, there is a great need for quantum computation using as few gates as possible. One way of achieving this is by using universal sets of gates (gate-sets) of high {\it efficiency}, i.e. such that can approximate any unitary with circuits of minimal length. This idea is also connected to complexity of unitaries (see \cite{susskind} for more details).

The efficiency of a universal set $\S$ (see \cite{sawicki2017,sawicki22017, sawicki2021} for criteria that allow deciding universality) is typically measured by the length of a circuit needed to approximate any quantum transformation with a given precision $\epsilon$.
The Solovay-Kitaev theorem states that all \textcolor{black}{symmetric} universal sets \footnote{A set $S\subset U(d)$ is symmetric if for any $U\in \mc{S}$ the inverse $U^{-1}\in \mc{S}$} are roughly the same efficient.
More precisely, the length of a circuit that $\epsilon$-approximates any $U\in SU(d)$ is bounded by $A(\S)\log^c(1/\epsilon)$ \cite{NC00}, where $c\geq 1$.
There have been recently some new developments connected to the Solovay-Kitaev theorem for gate-sets without inverses. 
First, in \cite{varju,oszmaniec} it was shown that an $\epsilon$-approximate  poly-log length circuit  exists even if one drops the assumption that a set $\mc{S}$ is symmetric.
Moreover in \cite{Bouland21} an algorithm implementing this sequence was given.
To estimate the value of $A(\S)$ one can use the concept of $\delta$-approximate $t$-designs \cite{harrow,oszmaniec}. 
To this end let  $\{\mathcal{S},\nu_{\mathcal{S}}\}$ be an ensemble of quantum gates, where $\mathcal{S}$ is a finite subset of $U(d)$ and $\nu_{\mathcal{S}}$ is a probability measure on $\mathcal{S}$. Such an ensemble is called 
 $\delta(\nu_{\mathcal{S}},t)$-approximate $t$-design if and only if 
\begin{equation*}
\delta(\nu_{\mathcal{S}},t)\coloneqq\left\|T_{\nu_{\mathcal{S}},t}-T_{\mu,t} \right\| < 1\  ,
\end{equation*}
where $\|\cdot\|$ is the operator norm and for any measure $\nu$ (in particular for the Haar measure $\mu$) we define a \emph{moment operator}
 \begin{gather*}
T_{\nu,t}\coloneqq\int_{U(d)} d\nu(U) U^{ t,t},\,\,\mathrm{where}\;
U^{t,t}=U^{\otimes t}\otimes \bar{U}^{\otimes t},
\end{gather*}
where $\bar{U}$ is entry-wise conjugation of $U$. When $\delta(\nu_\mc{S},t)=0$ we say that $\{\mc{S},\nu_\mc{S}\}$ is an {\it exact $t$-design}. Thus  (approximate) $t$-design are ensembles of unitaries that (approximately) recover Haar averages of polynomials in entries of unitaries up to the order $t$. More precisely any balanced polynomial of degree $t$ on $U(d)$ can be written as $f_{A}(U)=\tr\left(AU^{t,t}\right)$, where $A$ is a fixed matrix of size $d^{2t}\times d^{2t}$. Assuming that $\{\mathcal{S},\nu_\mc{S}\}$ is a $\delta$-approximate $t$-design we have
\begin{gather}\label{eq:approximate-polynomial}
\left|\int_{U(d)}d\nu_\mc{S}(U)f_A(U)-\int_{U(d)}d\mu(U) f_A(U)\right|=\left|\tr\left(A\left(T_{\nu_\mc{S},t}-T_\mu,t\right)\right)\right|\leq \|A\|_1\delta(\nu_\mc{S},t),
\end{gather}
where $\|A\|_1=\tr\sqrt{AA^\dagger}$. Thus $\delta(\nu_\mc{S},t)$ controls the error we make when taking average of $f_A$ with respect to $\nu_\mc{S}$ instead of the Haar measure. Unitaries constituting $\delta$-approximate $t$-design form $\epsilon$-nets  for    $t\simeq\frac{d^{5/2}}{\epsilon}$ and $\delta\simeq
\left(\frac{\epsilon^{3/2}}{d}\right)^{d^2}$ \cite{oszmaniec}. As a direct consequence of property \eqref{eq:approximate-polynomial} $\delta$-approximate $t$-designs find numerous applications throughout quantum information, including randomized benchmarking \cite{Gambetta2014,Brandao21}, information transmission \cite{InfTransmission2009}, quantum state discrimination \cite{StateDiscrimination2005}, criteria for universality of quantum gates \cite{sawicki2021} and complexity growth \cite{susskind,ChaosDesign2017,MO1,jonas2}. It is also known that the constant $A(\S)$ from the Solovay-Kitaev theorem is inversely proportional to  $1-\delta(\nu_\S)$, where $\delta(\nu_\S):=\mathrm{sup}_t\delta(\nu_\S,t)$, whenever $\delta(\nu_\S)<1$ \cite{harrow}.  Determining the value of $\delta(\nu_\S)$ requires computation of the norm of an infinite number of operators $T_{\nu_\S,t}$ which is analytically and numerically intractable. It is known, however, that  $\delta(\nu_\S)<1$ under the additional assumption that gates have algebraic entries \cite{bourgain2012, bourgain2008}. In this case also the constant $c$ in the Solovay-Kitaev theorem is equal to $1$. Recent results  \cite{bocharov,kliuchnikov,kliuchnikov2016, selinger, sarnak} based on some number theoretic constructions give examples of universal sets with $c = 1$ and the smallest possible value of $\delta(\nu_\S)$. The approach presented in these contributions has been unified in \cite{sarnak} where the author pointed out the connection of these new results to the seminal work concerning distributions of points on the sphere $S^2$ \cite{lubotzky}.  

In contrast to the above mentioned contributions, in this work, we do not focus on concrete gate-sets but instead we aim to answer the natural question of how likely it is that a set of gates has high efficiency. In order to achieve this goal we need to characterize efficiency of random universal gate-sets. Calculation of $\delta(\nu_\S)$ is of course intractable. Therefore we follow the approach of \cite{oszmaniec,varju} and consider $\delta(\nu_\S,t)$ for some fixed $t$ (which is determined by a precision $\epsilon$). The results of \cite{oszmaniec,varju} ensure that for a given precision $\epsilon$ the constant $A(\S)$ in the Solovay-Kitaev theorem is inversely proportional to $1-\delta(\nu_\S,t(\epsilon))$, where $t(\epsilon)=O(\epsilon^{-1})$,  and the constant $c=1$. Therefore, in order to characterize efficiency of random universal sets of gates we need to characterize a probability distribution of $\delta(\nu_\S,t)$.

What remains is to make precise what kind of random gate-sets we want to consider. As there is a natural uniform measure on the unitary group, i.e. the Haar measure one can consider two types of gate-sets:
\begin{enumerate}
    \item $\mathcal{S}=\{U_1,\ldots,U_n\}$, where $U_k$'s are independent and Haar random unitaries from $U(d)$. Such $\mc{S}$ will be called {\it Haar random  gate-set}.
    \item $\mathcal{S}=\{U_1,\ldots,U_n\}\cup\{U_1^{-1},\ldots,U_n^{-1}\}$, where $U_k$'s are independent and Haar random unitaries from $U(d)$. Such $\mc{S}$ will be called {\it symmetric Haar random  gate-set}.
\end{enumerate}

Another choice would be to start with an ensemble $\{\mathcal{D},\nu_{\mc{D}}\}$ that forms an exact $t$-design and consider two sets:
\begin{enumerate}
    \item $\mathcal{S}=\{U_1,\ldots,U_n\}\subset \mc{D}$, where $U_k$'s are independent and distributed according to $\nu_{\mc{D}}$. Such $\mc{S}$ will be called {\it t-random  gate-set}.
    \item $\mathcal{S}=\{U_1,\ldots,U_n\}\cup\{U_1^{-1},\ldots,U_n^{-1}\}$, where $\{U_1,\ldots,U_n\}\subset \mc{D}$ and $U_k$'s are independent and distributed according to $\nu_{\mc{D}}$. Such $\mc{S}$ will be called {\it symmetric $t$-random  gate-set}.
\end{enumerate}
We note that putting $\mathcal{D}=U(d)$ and $\nu_{\mc{D}}=\mu$, where $\mu$ is the Haar measure on $U(d)$, we get that a (symmetric) Haar random gate-set is a (symmetric) $t$-random gate set for any $t$. Thus (symmetric) Haar random gate-sets are (symmetric) $\infty$-random gate-sets and all the results that we prove for the (symmetric) $t$-random gate-sets are automatically true for (symmetric) Haar random gate-sets. 

In order to simplify the notation we will often denote the cardinality of $\mc{S}$ by $\mathcal{S}$ (instead of $|\mc{S}|$). Our main results are given in a form of bounds on the probability $\mathbb{P}\left(\delta(\nu_\S,t)\geq \delta \right)$, where $\mathcal{S}$ is a (symmetric) $t$-random gate set and $\nu_{\mc{S}}$ is a uniform measure on $\mc{S}$. In order to obtain them we first show that $T_{\nu_\S,t}$ decomposes as a direct sum over irreducible representations of the unitary group $U(d)$ that can be labeled by elements subset $\lambda \in \Lambda_t \subset \mathbb{Z}^d$ (see formula \eqref{eq:big-lambda}). The blocks appearing in this decomposition, that we denote by $T_{\nu_\S,\lambda}$, have dimensions $d_\lambda$ given by the Weyl dimension formula \eqref{eq:weyl_dimension}. 
Using the union bound 
\begin{gather}
\mathbb{P}\left(\delta(\nu_\S,t)\geq \delta \right)\leq \sum_{\lambda \in \Lambda_t} \mathbb{P}\left(\delta(\nu_\S,\lambda)\geq \delta \right),
\end{gather}
  we reduce the problem to finding bounds on $\mathbb{P}\left(\delta(\nu_\S,\lambda)\geq \delta \right)$. We achieve this combining recently derived matrix concentration inequalities \cite{tropp2015} with the recent result concerning calculation of higher Frobenius-Schur indicators for semisimple Lie algebras \cite{abuhamed2007}, that we further improve. Our main results are Theorems \ref{th:master_bound-main1}, \ref{th:master_bound-main2} that give concrete calculable bounds on $\mathbb{P}\left(\delta(\nu_\S,t)\geq \delta \right)$.

\begin{theorem}
\label{th:master_bound-main1}
Let $\mathcal{S}$ be a $t$-random gate-set and $\nu_\mc{S}$ a uniform measure. Then for any $\delta < 1$
\begin{gather}
    \mathbb{P}\left(\delta(\nu_\S,t)\geq \delta\right)\leq  \frac{2e^{-\delta\mc{S}\,\mathrm{arctanh}(\delta)}}{(1-\delta^2)^{\frac{\mc{S}}{2}}}\sum_{\lambda\in \Lambda_t} d_\lambda,
    \end{gather}
 where, $\Lambda_t$ is given by \eqref{eq:big-lambda} and $d_\lambda$ is given by \eqref{eq:weyl_dimension}.
\end{theorem}

\begin{theorem}
\label{th:master_bound-main2}
Let $\mathcal{S}$ be a symmetric Haar random gate-set. Then for any $\delta < 1$
\begin{gather}
    \mathbb{P}\left(\delta(\nu_\S,t)\geq \delta\right) \leq \sum_{\lambda\in\Lambda_t} d_\lambda e^{-\frac{\mc{S}\delta^2}{\sqrt{1-\delta^2}}} \left[ F\left(\frac{\mc{S}\delta}{\sqrt{1-\delta^2}},\lambda, \mc{S}\right) +  F\left(-\frac{\mc{S}\delta}{\sqrt{1-\delta^2}},\lambda, \mc{S}\right) \right], 
    \end{gather}
where $\Lambda_t$ is given by \eqref{eq:big-lambda},  $d_\lambda$ is given by \eqref{eq:weyl_dimension} and $F(\cdot,\cdot,\cdot)$ is given by \eqref{eq:F}.
    \end{theorem}

These theorems are then used to obtain bounds on the size of a \textcolor{black}{$t$-}random gate-set needed to form, with a probability $P$, a $\delta$-approximate $t$-designs, for various $t$'s and $\delta$'s. We show that this size is of the order $O( \delta^{-2}(t\log(d)-\log(1-P)))$. Moreover, we compare the number of independent $t$-random $n$-qubit gates, $\mc{S}_n$, needed to form $0.01$-approximate $2$-design with the probability $0.99$ and the size of the $n$-qubit Clifford group, $\mc{C}_n$, that is known to be an exact $2$-design. The ratio of $\mc{C}_n/\mc{S}_n$ turns out to grow exponentially with the number of qubits.\\
In Section \ref{subsec:bounds-for-circuits} we also show that Theorems \ref{th:master_bound-main1} and \ref{th:master_bound-main2} can be easily generalised to a scenario where instead of $t$-random set of gates we have a set of random quantum circuits composed of $t$-random independent gates.

Theorem \ref{th:master_bound-main1} can be used when one needs to calculate average of any $(t,t)$-polynomial over $U(d)$. Such polynomials arise, for example, in the randomized benchmarking protocols where one is interested in assessing quality of quantum gates implementation and considers the $k$-th moments of the fidelity \cite{Emerson_2005,Dankert09,Yoshifumi2}.
\begin{gather}\label{eq:average-fidelity}
    \int_{U(d)}d\mu(U)\tr\left(\rho U^{-1}\Phi(U\rho U^{-1})U\right)^k,
\end{gather}
where $\Phi$ is a quantum channel that represents gate independent noise and for perfect implementation of $U$ is the identity. Of course one can  replace the Haar measure in \eqref{eq:average-fidelity} by an exact $2k$-design as the integrated function is $(2k,2k)$-polynomial. Using our results we can replace an exact $2k$-design, which as shown in \cite{Yoshifumi2} has exponential size in $n\sqrt{2k}$, where  $n$ is a number of qubits, by a $\delta$-approximate $t$-designs of size (see Section \ref{sec:inequalities_comparison})
\begin{gather*}
\mc{S}\geq \frac{2\left(2t\log\left(d\right)+\log(2)-\log\left(1-P\right)\right)}{\log\left(\left(1+\delta\right)^{1+\delta}\left(1-\delta\right)^{1-\delta}\right)}.
\end{gather*}
Moreover we can control the error using  \eqref{eq:approximate-polynomial}.

    
Finally we analyze concentration properties of $\delta(\nu_\S,t)$ around its mean value $\E{\mu} \delta(\nu_\S,t)$. Our main result is 
\begin{theorem}\label{th:intro_master_bound-main3}
Let $\mathcal{S}$ be a Haar random  gate-set. Then
\begin{gather}
\mathbb{P}\left(\delta(\nu_{\mathcal{S}},t)\geq \E{\mu} \delta(\nu_{\mathcal{S}},t)+\alpha\right)\leq \exp\left({\frac{-d\mathcal{S}\alpha^2}{32t^2}}\right).
\end{gather}
\end{theorem}
The methods behind its proof \cite{meckes} can be extended to Haar random gate-sets with particular structure or architecture. Following this observation we analyze efficiency of random $d$-mode circuits build from $2$-mode beamsplitters. More precisely we consider the Hilbert space $\mathcal{H}=\mc{H}_1\oplus\ldots\oplus\mc{H}_d$, where $\mc{H}_k\simeq\mathbb{C}$, $d>2$ and we call spaces $\mc{H}_k$ modes. For a matrix $B\in SU(2)$, which we call a $2$-mode beamsplitter, we define matrices $B^{ij}$, $i\neq j$, to be the matrices that act on a $2$-dimensional subspace $\mc{H}_i\oplus\mc{H}_j\subset \mc{H}$ as $B$ and on the other components of $\mc{H}$ as the identity. This way a matrix $B\in SU(2)$ gives $d(d-1)$ matrices in $SU(d)$. Applying this procedure to a Haar random gate-set set $S\subset SU(2)$ we obtain random gate-set $S^d$ (see \cite{Reck,sawicki1}) and it is natural to ask about its efficiency. Our main conclusion here is that a Haar random gate-set $\mc{S}\subset SU(2)$ gives the gate-set $S^d\subset SU(d)$ for which $\delta(\nu_{\mathcal{S}^d},t)$ has the same concentration rate around the mean as a Haar random gate-set $\mc{S}^\prime\subset SU(d)$ of size: $\mc{S}^\prime=\frac{2\mc{S}}{d}$, i.e. 
 \begin{theorem}\label{th:intro_beam}
Let $\mc{S}\subset SU(2)$ be a Haar random gate-set and $\mc{S}^d\subset SU(d)$ the corresponding $d$-mode gate-set. Then 
\begin{gather}
\mathbb{P}\left(\delta(\nu_{\mathcal{S}^d},t)\geq \E{\mu}\delta(\nu_{\mathcal{S}^d},t)+\alpha\right)\leq \exp\left({\frac{-\mathcal{S}\alpha^2}{16t^2}}\right).
\end{gather}
\end{theorem}
Using similar methods one can show the concentration around the mean value of any function 
\begin{equation*}
    F: U(d)^{\times n} \ni (U_1, ..., U_n) \rightarrow F(U_1, ..., U_n) \in \mathbb{R},
\end{equation*}
for $U_k$'s independent and Haar random as long as $F$ is $L$-Lipschitz:
\begin{equation*}
    | F(U_1, ..., U_n) - | F(V_1, ..., V_n) | \leq L\left(\sum_{k=1}^n \|U_k-V_k\|^2_{F}\right)^{\frac{1}{2}},
\end{equation*}
where $\| \cdot \|_F$ is a Frobenius (Hilbert-Schmidt) norm. We explain this in detail in Section \ref{sec:inequalities_review}. A function $F$ can be, for example, given by the $k$-th moment of the fidelity of a quantum circuit of the length $n$ (see \cite{Yoshifumi2} for more details).

\color{black}

The paper is organized as follows. In Section \ref{sec:inequalities_review} we present a short review of matrix concentration inequalities that will play a central role in our setting. Next, in Section \ref{sec:irreps} we provide necessary information concerning irreducible representations of unitary groups. Then in Section \ref{sec:moments} we introduce notion of moment operators. {\color{black} In Section \ref{sec:fs-indicators} we explain the role of Frobenius-Schur indicators and how to compute them.} The main results of these paper are then showed in Section \ref{sec:main_results}, while Section \ref{sec:inequalities_comparison} contains analysis of the results and applications.

\section{Short review of relevant matrix concentration inequalities }\label{sec:inequalities_review}
In this section we review known upper bounds on the probability that $\mathbb{P}(F(X)\geq \delta )$, for classes of random matrices $X$ and real valued functions $F$ that are relevant in our setting. An interested reader is referred to \cite{tropp2015, meckes} for more details. The first class of inequalities concerns a situation when $X=\sum X_k$, where $X_k\in \mathrm{Mat}(d,\mathbb{C})$ are independent, random, Hermitian matrices and the function $F$ is the operator norm of $X$, $F(X)=\|X\|$. Thus we are looking for an upper bound for $\mathbb{P}\left(\|X\|\geq \delta\right)$. The line of reasoning is as follows. Let $\lambda_{\mathrm{max}}(X)$ and $\lambda_{\mathrm{min}}(X)$  denote the largest and the smallest eigenvalues of $X$ respectively. In the first step one uses the exponential Markov inequality, i.e. 

\begin{gather*}
    \mathbb{P}\left(Y\geq t\right)=\mathbb{P}\left(e^{\theta Y}\geq e^{\theta t}\right)\leq e^{-\theta t}\mathbb{E}e^{\theta Y},
\end{gather*}
for any $\theta >0$. Taking $Y=\lambda_{\mathrm{max}}(X)$ and using the fact that $e^{\theta \lambda_{\mathrm{max}}(X)}\leq \mathrm{tr} e^{\theta X}$ we get

\begin{gather}\label{expMarkov}
    \mathbb{P}\left(\lambda_{\mathrm{max}}(X)\geq \delta\right)\leq \mathrm{inf}_{\theta >0}e^{-\theta \delta}\mathbb{E}\mathrm{tr}e^{\theta X}.
    \end{gather}
Next we note that $\mathbb{P}\left(\lambda_{\mathrm{min}}(X)\leq \delta\right)= \mathbb{P}\left(-\lambda_{\mathrm{min}}(X)\geq -\delta\right)= \mathbb{P}\left(\lambda_{\mathrm{max}}(-X)\geq -\delta\right)$. Thus     
    
 \begin{gather}\label{lambdamin}
  \mathbb{P}\left(\lambda_{\mathrm{min}}(X)\leq \delta\right)\leq \mathrm{inf}_{\theta >0}e^{\theta \delta}\mathbb{E}\mathrm{tr}e^{-\theta X}.
\end{gather}
Next, using the Lieb theorem \cite{Lieb} (one can alternatively use the Golden-Thomson inequality \cite{Golden,thompson} but the resulting bound is in general weaker \cite{tropp2015} ) we obtain 
\begin{gather}\label{Lieb}
\mathbb{E}\mathrm{tr}e^{\theta \sum_k X_k}\leq \mathrm{tr}\exp\left(\sum_k\log\mathbb{E}e^{\theta X_k}\right),  
\end{gather}
for any $\theta \in\mathbb{R}$. Combining \eqref{expMarkov} and \eqref{lambdamin} with \eqref{Lieb} we get 
\begin{gather}\label{lambdaminmax}
    \mathbb{P} \left( \lambda_{\mathrm{max}}(X) \geq \delta \right) \leq \inf_{\theta>0} e^{-\theta \delta} \mathrm{tr} \exp \left( \sum_k \log \mathbb{E} e^{\theta X_k} \right),\\
    \mathbb{P} \left( \lambda_{\mathrm{min}}(X)\leq \delta \right) \leq  \inf_{\theta>0} e^{\theta \delta} \mathrm{tr} \exp \left( \sum_k \log \mathbb{E} e^{-\theta X_k} \right).
\end{gather}
Finally,
\begin{gather*}
    \mathbb{P} \left( \|X\| \geq \delta \right) = \mathbb{P} \left( \mathrm{max}\{\lambda_{\mathrm{max}}(X),-\lambda_{\mathrm{min}}(X)\} \geq \delta \right) \leq\\\leq \mathbb{P} \left( \lambda_{\mathrm{max}}(X) \geq \delta \right)+\mathbb{P} \left( \lambda_{\mathrm{min}}(X)\leq -\delta \right).
\end{gather*}

\begin{fact}[Master bound]
\label{fact:master_bound}
Let $X=\sum X_k$, where $X_k\in \mathrm{Mat}(d,\mathbb{C})$ are independent, random, Hermitian matrices. Then
\begin{gather*}
    \mathbb{P} \left( \|X\| \geq \delta \right) \leq  \inf_{\theta>0} e^{-\theta \delta} \mathrm{tr} \exp \left( \sum_k \log \mathbb{E} e^{\theta X_k}\right)+\inf_{\theta>0} e^{-\theta \delta} \mathrm{tr} \exp \left( \sum_k \log \mathbb{E} e^{-\theta X_k} \right).
\end{gather*}
\end{fact}

A master bound can be also derived for the sum of non-Hermitian random matrices. To this end, following \cite{tropp2015}, we consider the Hermitian dilation map 
\begin{gather*}
    \mathcal{H}:\mathrm{Mat}(d,\mathbb{C})\rightarrow \mathbb{H}(2d),
\end{gather*}
where $\mathbb{H}(2d)$ is the space of $2d\times 2d$ Hermitian matrices given by
\begin{gather*}
\label{eq:hermitization}
    \mathcal{H}(X)=\begin{pmatrix}
0& X\\
X^\dagger &0
\end{pmatrix}.
\end{gather*}
This map is clearly a real linear map. One can also show (cf. \cite{tropp2015}) that $\|X\|=\|\mathcal{H}(X)\|=\lambda_\mathrm{max}\left(\mathcal{H}(X)\right)$. Making use of \eqref{lambdaminmax} we get:

\begin{fact}\label{fact:master-nonhermitian}
Let $X=\sum X_k$, where $X_k\in \mathrm{Mat}(d,\mathbb{C})$ are independent random matrices. Then 
\begin{gather*}
    \mathbb{P}\left(\|X\|\geq \delta\right)\leq \inf_{\theta>0}e^{-\theta\delta}\mathrm{tr}\exp\left(\sum_k\log\mathbb{E}e^{\theta\mathcal{H}(X_k)}\right).
\end{gather*}
\end{fact}
The upper bounds in Facts \ref{fact:master_bound} and \ref{fact:master-nonhermitian} can be further simplified by finding a majorization of $\log \mathbb{E}e^{\theta X_k}$ in terms of moments $\mathbb{E}X_k^n$  that allow analytic optimization over $
\theta$ (see chapter 6 of \cite{tropp2015} for more details). Usage of moments up to order two leads to the matrix Bernstein inequality.  
\begin{fact}[Matrix Bernstein inequality]
\label{fact:bernstein}
Let $X=\sum X_k$, where $X_k\in \mathrm{Mat}(d,\mathbb{C})$ are independent, random matrices such that:
\begin{equation*}
    \forall_k \quad \mathbb{E}X_k = 0 \quad \mathrm{and} \quad \| X_k \| \le L.
\end{equation*}
Let $v$ be the matrix variance statistic of the sum:
\begin{align*}
    v &= \max \left\{ \| \mathbb{E}(XX^\dagger) \|,\, \| \mathbb{E}(X^\dagger X) \| \right\}, \\
    &= \max \left\{ \| \sum_k \mathbb{E}(X_k X_k^\dagger) \|,\, \| \sum_k \mathbb{E}(X_k^\dagger X_k) \| \right\}.
\end{align*}
Then for all $\delta \ge 0$
\begin{equation*}
    \mathbb{P}\left(\| X \| \ge \delta \right) \le 2 d \exp \left( \frac{-\delta^2/2}{v + L \delta/3} \right).
\end{equation*}
\end{fact}
\subsection{Bounds for Haar random matrices}
The second type of inequalities we will consider in this paper \textcolor{black}{rely} on the fact that the randomness comes from the Haar measure. The interested reader is \textcolor{black}{referred} to chapter 5 of  \cite{meckes} for detailed discussion. Here we only give the main result and mention that its proof is based on the fact that, by Bakery-\'Emery curvature criterion, the (special)unitary group equipped with the Haar measure satisfies the so-called logarithmic Sobolev inequality which, via the Herbst argument, leads to concentration of measure for Lipschitz functions.
\begin{fact}\label{fact:haar_inequality}
Let $G^{\times \mathcal{S}}$, where $G$ is $SU(d)$ or $U(d)$, be equipped with the metric given by the $L_2$-sum of Hilbert-Schmidt (Frobenius) metrics on the group $G$, i.e. the distance between $(U_1,\ldots,U_\mc{S})\in G^{\times \mc{S}}$ and $(V_1,\ldots,V_\mc{S})\in G^{\times \mc{S}}$ is 
\begin{gather*}
\left(\sum_{k=1}^\mc{S}\|U_k-V_k\|^2_{F}\right)^{\frac{1}{2}} .
\end{gather*}
Suppose that
\begin{equation*}
    F: G^{\times \mathcal{S}} \ni (U_1, ... , U_\mc{S}) \rightarrow F(U_1, ... , U_\mc{S}) \in \mathbb{R},
\end{equation*}
is $L$-Lipschitz, i.e.
\begin{gather*}
    |F(U_1, ... , U_\mc{S})-F(V_1, ... , V_\mc{S})|\leq L\left(\sum_{k=1}^\mc{S}\|U_k-V_k\|^2_{F}\right)^{\frac{1}{2}} ,
\end{gather*}
and that $U_1, ..., U_\mc{S}$ are independent and chosen according to the Haar measure on $G$. Then for any $\alpha > 0$
\begin{equation*}
    \mathbb{P}\left( F(U_1, ... , U_\mc{S}) \ge \mathbb{E}F(U_1, ... , U_\mc{S}) + \alpha \right) \le \exp \left( -\frac{d\alpha^2}{4CL^2} \right).
\end{equation*}
where $C$ is equal to $2$ for $SU(d)$ and $6$ for $U(d)$.
\end{fact}

\section{Irreducible representations of $U(d)$ and $SU(d)$} \label{sec:irreps}
In this section we  recall some basic facts about Lie groups, Lie algebras and their representations in the context of groups $G=U(d)$ and $G=SU(d)$. In Table \ref{tab:lie} we summarize information about those groups, where we used $M_d^0(\CC) := \left\{ X \in M_d(\CC) \, | \, \mathrm{Tr}(X) = 0 \right\}$.
We will denote by $\mk{g}$ the Lie algebra of $G$. 
The Lie algebra $\mk{su}(d)$ has no non-trivial ideals and thus is semisimple. On the other hand the algebra $\mk{u}(d)$ has an abelian ideal consisting of matrices proportional to the identity and is not semisimple. We note, however, that $[\mk{u}(d),\mk{u}(d)]=\mk{su}(d)$. We will call a Lie group semisimple if its Lie algebra is semisimple. Other relevant for us algebras are Lie algebra complexification $\mathfrak{g}_\mathbb{C} := \mathfrak{g} + i\mathfrak{g}$, the Lie algebra of the maximal torus $T$ in $G$ - $\mathfrak{t}$  and the Cartan subalgebra (CSA) - $\mathfrak{h} := \mk{t} + i\mk{t}$.   
\begin{table}[htp]
    \centering
    \begin{tabular}{c|c|c}
         $G$ & $U(d)$ & $SU(d)$ \\
         $\mk{g}$ & $\mk{u}(d) := \left\{X \in  M_d(\CC) \, | \, X^\dagger = -X\right\}$ & $\mk{su}(d) := \mk{u}(d) \cap M_d^0(\CC)$ \\
         semi-simple & no & yes \\
         $\mk{g}_\CC$ & $\mk{u}(d)_\CC \cong \mk{gl}(d, \CC) := M_d(\CC) $ & $\mk{su}(d)_\CC \cong \mk{sl}(d, \CC) := M_d^0(\CC)$ \\
         $\mk{t}$ & $\mk{t} := \left\{ X \in \mk{u}(d) \, | \, X\,\mathrm{diagonal} \right\}$ & $\mk{t}_0 := \mk{t} \cap M_d^0(\CC)$ \\
         $\mk{h}$ & $\mk{h} := \left\{ X \in \mk{gl}(d, \CC) \, | \, X\,\mathrm{diagonal} \right\}$ & $\mk{h}_0 := \mk{h} \cap M_d^0(\CC)$
    \end{tabular}
    \caption{The comparison of groups $U(d)$ and $SU(d)$ in view of the Lie group theory. We used $M_d^0(\CC)$ to denote $\left\{ X \in M_d(\CC) \, | \, \mathrm{Tr}(X) = 0 \right\}$.}
    \label{tab:lie}
\end{table}

The functional $\alpha \in \mk{h}^*$ is called a {\it root} of $\mk{g}$ if and only if there exists $X_\alpha \in \mk{g}_\CC$ such that:
\begin{equation}
    \label{eq:root}
    \forall_{H\in\mk{h}}\; \left[H, X_\alpha\right] = \alpha\left(H\right)X_\alpha.
\end{equation}
We denote the set of all roots by $\Phi$ and call it {\it the root system}. For a given root $\alpha$ the subspace of all $X_\alpha$ satisfying \eqref{eq:root} is called a root subspace of $\alpha$ and denoted by $\mk{g}_\alpha$. The algebra $\mk{g}_\CC$ decomposes
\begin{equation*}
    \mk{g}_\CC = \mk{h} \oplus \bigoplus_{\alpha \in \Phi} \mk{g}_\alpha.
\end{equation*}
If we define $L_i, \alpha_{i, j} \in \mk{h}^*$ as
\begin{gather}
    \label{eq:L}
    L_i \left( 
    \begin{pmatrix} 
          a_1 & 0 & 0 \\
          0 & \ddots & 0 \\
          0 & 0 & a_d
    \end{pmatrix} 
    \right) := a_i,\\
    \alpha_{i, j} := L_i - L_j,
\end{gather}
then the root system for $U(d)$ and $SU(d)$ is
\begin{equation*}
    \Phi = \left\{ \alpha_{i, j} \, | \, 1 \le i, j \le d \right\}.
\end{equation*}
Among roots we distinguish positive roots
\begin{equation*}
    \Phi^+ := \left\{ \alpha_{i, j} \, | \, 1 \le i < j \le d \right\},
\end{equation*}
and positive simple roots
\begin{equation*}
    \Delta := \left\{ \alpha_{i, i+1} \, | \, 1 \le i  \le d-1 \right\}.
\end{equation*}
For $\alpha, \beta \in \mk{h}^*$ we say that $\alpha$ is higher (lower) than $\beta$ iff $\alpha - \beta$ is a linear combination of positive simple roots with non-negative (non-positive) coefficients and we denote it by $\alpha > \beta$ ( $\alpha < \beta$).\\
We introduce the inner product on $\mk{h}^*$ defined by
\begin{equation*}
    \braket{L_i|L_j} := \delta_{i, j}.
\end{equation*}
The inner product gives us an isomorphism $\mk{h}^* \ni L \mapsto \bra{\bar{L}} \in \mk{h}$ hence further in the text we will identify $\mk{h}^*$ with $\mk{h}$. For any $\alpha \in \Phi$ the root system is preserved under the reflection about the hyper-plane perpendicular to $\alpha$:
\begin{equation*}
    s_\alpha : \mk{h} \ni H \mapsto H - \frac{2\braket{H|\alpha}}{\braket{\alpha|\alpha}}\alpha.
\end{equation*}
The group $\mc{W} := <s_\alpha \,|\, \alpha \in \Phi >$ is called the Weyl group of $\Phi$. In our case $\mc{W}$ is isomorphic to the group of permutations $S_d$ and action of $\sigma \in \mc{W}$ on $\mk{h}$ is given by:
\begin{equation*}
    \sigma \cdot L_i := L_{\sigma^{-1}(i)}.
\end{equation*}
An element $H \in \mk{h}$ is called:
\begin{itemize}
    \item integral iff $\forall_{\alpha \in \Phi} \; \frac{2\braket{H|\alpha}}{\braket{\alpha|\alpha}} \in \mathbb{Z}$,
    \item analytically integral iff for all $X \in \mk{h}$ such that $e^{2\pi i X} = \bbone$ there holds $\braket{X|H} \in \mathbb{Z}$,
    \item dominant iff $\forall_{\alpha\in\Delta}\; \braket{\alpha|H} \ge 0$.
\end{itemize}
For every finite dimensional representation $\pi: G \rightarrow GL(V)$ of $G$ there exists representation $\rho: \mk{g} \rightarrow \mk{gl}(V)$ such that for any $X\in\mk{g}$ there holds
\begin{equation*}
    \pi\left(e^X\right) = e^{\rho(X)}.
\end{equation*}
Let us define the complexification of $\rho$ to be $\rho_\CC: \mk{g}_\CC \ni X+iY \mapsto \rho(x) + i \rho(Y) \in \mk{gl}(V, \CC)$. Then the following are equivalent:
\begin{itemize}
    \item $\pi$ is irreducible,
    \item $\rho$ is irreducible, 
    \item $\rho_\CC$ is irreducible.
\end{itemize}
Further in the text we will slightly abuse notation and we will use $\rho$ also for $\rho_\CC$. A weight of $\rho$ is an integral element $\mu$ such that there exists a non-zero vector $v_\mu \in \rho$ satisfying:
\begin{equation}
    \label{eq:weight_vector}
    \forall_{H\in \mk{h}}\; \rho(H)v_\mu = \braket{\mu|H}v_\mu.
\end{equation}
Subspace of all $v_\mu$ satisfying \eqref{eq:weight_vector} is called the weight space of $\mu$ and we denote it by $\rho^\mu$ or $\pi^\mu$. The multiplicity $m^\mu$ (or $m(\mu)$) of $\mu$ is the dimension of its weight space. Every irreducible representation is a direct sum of its weight spaces. The notions of weight space and root space are closely connected. Indeed for $v_\mu \in \rho^\mu$, $X_\alpha \in \mk{g}_\alpha$ and $H \in \mk{h}$ we have:
\begin{gather*}
    \rho\left(H\right)\rho\left(X_\alpha\right)v_\mu = \left( \rho(\left[H, X_\alpha\right]) + \rho\left(X_\alpha\right)\rho\left(H\right)\right)v_\mu = \\ =
    \left( \braket{\alpha|H} + \braket{\mu|H} \right) \rho\left(X_\alpha\right)v_\mu = \braket{\mu + \alpha | H} \rho\left(X_\alpha\right)v_\mu,
\end{gather*}
thus $\rho\left(X_\alpha\right)v_\mu$ is either $0$ or in $\rho^{\mu+\alpha}$. The highest weight $\lambda$ is a weight in $\rho$ that is higher than all other weights in $\rho$. Now, we can state the theorem of the highest weight.
\begin{fact}[Theorem of the highest weight]
\label{th:highest_weight}
For semi-simple complex Lie algebra $\mk{g}_\CC$ and its finite-dimensional representation $\rho$ we have:
\begin{enumerate}
    \item $\rho$ has unique highest weight $\lambda$,
    \item $\lambda$ is dominant,
    \item if $\tilde{\rho}$ is another representation of $\mk{g}_\CC$ with highest weight $\lambda$ then $\tilde{\rho}$ is isomorphic to $\rho$,
    \item if $\lambda$ is dominant and integral then there exists finite-dimensional irreducible representation of $\mk{g}_\CC$ with highest weight $\lambda$.
\end{enumerate}
\end{fact}
\begin{fact}
If $G$ is compact and connected the analogous theorem holds with the only difference that the highest weight $\lambda$ has to be analytically integral.
\end{fact}
In case of $U(d)$ we will identify highest weights $\lambda$ with sequences $(\lambda_1, ..., \lambda_d)$ where $\lambda_i := \braket{\lambda|L_i}$ and in case of $SU(d)$ we will identify highest weights $\lambda^s$ with sequences $(\lambda^s_1, ..., \lambda^s_{d-1})$ where $\lambda^s_i := \braket{\lambda^s|\alpha_{i, i+1}}$. From the second point of Fact \ref{th:highest_weight} we have that for $i < j$ there holds $\lambda_i \ge \lambda_j$ and from analytical integrality $\lambda_i \in \mathbb{Z}$. Moreover, from the fourth point of Fact \ref{th:highest_weight} every sequence from $\mathbb{Z}^d$ satisfying those conditions uniquely defines the highest weight $\lambda$ and the associated representation. Analogously for $SU(d)$ the condition is $\lambda^s_i \in \mathbb{Z}_+$ and any element of $\mathbb{Z}^{d-1}_+$ defines the highest weight. In both cases we use $\pi_\lambda$ and $\rho_\lambda$ to denote the irreducible finite-dimensional representations with highest weight $\lambda$.

Since $SU(d)$ is a subgroup of $U(d)$ any representation $\pi_\lambda$ of $U(d)$ can be restricted to a representation $\pi'_{\lambda^s} := \pi_\lambda|_{SU(d)}$ of $SU(d)$. The question arises what is the relation between $\lambda$ and $\lambda^s$. On the other hand, irreducible representations of $SU(d)$ are often labeled by the Young diagrams instead of highest weights. In the following lemma we explore the relations between highest weights of $U(d)$, highest weights of $SU(d)$ and Young diagrams.
\begin{lemma}
\label{lemma:young_diagrams}
Let $\lambda$ be a highest weight of representation $\pi_\lambda$ of $U(d)$. Then the corresponding representation of $SU(d)$ has a Young diagram given by $(\lambda_1-\lambda_d,\ldots,\lambda_{d-1}-\lambda_{d})$ and the standard highest weight of this representation is given by $(\lambda_1-\lambda_2,\ldots,\lambda_{d-2}-\lambda_{d-1},\lambda_{d-1}-\lambda_{d})$.
\end{lemma}
\begin{proof}
By the definition of $\lambda$ we have that for the highest weight vector $v_\lambda \in \rho_\lambda^\lambda$ and any $H \in \mathfrak{h}$ it holds:
\begin{equation*}
    \rho_\lambda(H)v = \braket{\lambda|H}v_\lambda.
\end{equation*}
Then for the representation $\rho'_\lambda := \rho_\lambda|_{\mathfrak{sl}(d, \CC)}$ and its any subspace $V$ we have:
\begin{gather*}
    \forall_{X \in \mathfrak{su}(d)}\, \rho_\lambda'(X)V \subset V \Rightarrow \forall_{ X \in \mathfrak{su}(d),\; \phi \in \mathbb{R}}\, (\rho_\lambda'(X) + i\phi\bbone)V \subset V \Rightarrow \\
    \Rightarrow \forall_{ X \in \mathfrak{su}(d),\; \phi \in \mathbb{R}}\, 
    \rho_\lambda(X + i\phi\bbone)V \subset V \Rightarrow \forall_{X \in \mathfrak{u}(d)}\, \rho_\lambda(X)V \subset V.
\end{gather*}
Hence irreducibility of $\rho_\lambda$ implies the irreducibility of $\rho'_\lambda$.\\
Note that $L_1-L_d, ..., L_{d-1}-L_d$ is a basis of $\mathfrak{h}_0$ and for $i \in \{1, ..., d-1\}$:
\begin{equation*}
    \rho'_\lambda(L_i-L_d)v_\lambda = \rho_\lambda(L_i-L_d)v_\lambda =
    \braket{\lambda | L_i-L_d} v_\lambda = (\lambda_i - \lambda_d)v_\lambda.
\end{equation*}
Since $v_\lambda$ is the highest weight vector the sequence $\lambda^Y := (\lambda_1-\lambda_d, ..., \lambda_{d-1}-\lambda_d)$ uniquely determines the $\mathfrak{sl}(d, \CC)$ representation and the associated $SU(d)$ representation. Moreover, from the relations $\lambda_1 \ge ... \ge \lambda_d$ we have that $\lambda^Y$ is a sequence of non-negative, descending integers and as such can be interpreted as a Young diagram. If we choose $L_1-L_2, ..., L_{d-1}-L_d$ as a basis of $\mathfrak{h}_0$, we analogously obtain the sequence $\lambda^s := (\lambda_1-\lambda_2, ..., \lambda_{d-1}-\lambda_d)$ that is a highest weight of $SU(d)$ representation.
\end{proof}
To simplify the description of highest weights we introduce the following notation. For any $\lambda=(\lambda_1,\ldots,\lambda_d)$ in $\mathbb{Z}^d$ we will denote its length by $l(\lambda)=d$. By $\lambda_{+}$ we will denote the subsequence of $\lambda$ of positive integers. Moreover 
\begin{gather}
   \Sigma(\lambda)\coloneqq\sum_{k=1}^d\lambda_k,\\
   \label{eq:lamb-norm}
   \|\lambda\|_1\coloneqq\sum_{k=1}^d|\lambda_k|,\\
   d_\lambda \coloneqq \mathrm{dim}\pi_\lambda
\end{gather}
The value of $d_\lambda$ is determined by the Weyl dimension formula 
\begin{gather}
    \label{eq:weyl_dimension}
    d_\lambda=\prod_{1\leq i<j\leq d}\frac{\lambda_i-\lambda_j+j-i}{j-i}.
\end{gather}

\begin{fact}
\label{fact:possible_weights}
\cite{hall2004} Suppose that $\pi_\lambda$ is an irreducible representation of semi-simple Lie group $G$ then  $\mu \in \mk{h}$ is a weight of $\pi_\lambda$ if and only if the following two conditions are satisfied:
\begin{enumerate}
    \item $\mu$ is contained in the convex hull of the orbit of $\lambda$ under the Weyl group,
    \item $\lambda - \mu$ is a linear combination of positive simple roots with integer coefficients.
\end{enumerate}
\end{fact}
\begin{corollary}
Since $SU(d)$ is semi-simple and $SU(d)$ and $U(d)$ have the same roots and root spaces the above fact also applies to $U(d)$.
\end{corollary}

We use Fact \ref{fact:possible_weights} to prove the following lemma.

\begin{lemma}
\label{weight}
For any weight $\mu$ of the representation $\pi_\lambda$ we have $\|\mu\|_{1}\leq\|\lambda\|_1$ and $\Sigma(\mu)=\Sigma(\lambda)$.
\end{lemma}
\begin{proof}
The first condition of Fact \ref{fact:possible_weights} reads
\begin{equation*}
    \exists \; \{ t_\sigma \in [0, 1] \; | \; \sigma \in S_d \} \; \textrm{such that} \; \sum_{\sigma\in S_d}t_\sigma = 1 \; \textrm{and} \; \mu = \sum_{\sigma\in S_d}t_\sigma \sigma \cdot \lambda,
\end{equation*}
hence
\begin{equation*}
    \| \mu \|_1 = 
    \sum_{i=1}^d |\sum_{\sigma\in S_d}t_\sigma \lambda_{\sigma(i)}| \le 
    \sum_{i=1}^d \sum_{\sigma\in S_d}t_\sigma |\lambda_{\sigma(i)}| = 
    \sum_{\sigma\in S_d}t_\sigma \| \lambda \|_1 =
    \| \lambda \|_1,
\end{equation*}
which proves the first part of the lemma. The second condition from Fact \ref{fact:possible_weights} implies
\begin{equation*}
    \lambda - \mu = \sum_{i=1}^{d-1} c_i(\underbrace{0, ..., 0}_{i-1}, 1, -1, \underbrace{0, ..., 0}_{d - i -1}).
\end{equation*}
By acting $\Sigma$ on both sides we obtain
\begin{equation*}
    \Sigma(\mu - \lambda) = 0 \quad\Rightarrow\quad \Sigma(\mu) = \Sigma(\lambda).
\end{equation*}
\end{proof}

\begin{fact}[Kostant formula]
\label{fact:kostant}
Let $p: \mk{h}_0 \rightarrow \mathbb{N}$ be such that $p(\mu)$ is equal to the number of ways $\mu$ can be expressed as a linear combinations of positive simple roots with non-negative integer coefficients, $m_{\lambda^s}(\mu)$ a multiplicity of $\mu$ in $\pi_{\lambda^s}$ and $\rho$ the half-sum of positive roots\footnote{Note that the corresponding Lie algebra representation is denoted by $\rho_{\lambda^s}$} then
\begin{equation}\label{eq:Kostant}
    m_{\lambda^s}(\mu) = \sum_{\sigma \in S_d} \mathrm{sgn}(\sigma)\, p(\sigma \cdot (\lambda^s + \rho) - (\mu + \rho))
\end{equation}
\end{fact}

\section{Moment operators }\label{sec:moments}
Let  $\{\mathcal{S},\nu_{\mathcal{S}}\}$ be an ensemble of quantum gates, where $\mathcal{S}$ is a subset of $U(d)$ and $\nu_{\mathcal{S}}$ is \textcolor{black}{a probability} measure on $\mathcal{S}$. Such an ensemble is called 
 $\delta(\nu_{\mathcal{S}},t)$-approximate $t$-design if and only if 
\begin{equation*}
\delta(\nu_{\mathcal{S}},t)\coloneqq\left\|T_{\nu_{\mathcal{S}},t}-T_{\mu,t} \right\| < 1\  ,
\end{equation*}
where $\|\cdot\|$ is the operator norm and for any measure $\nu$ (in particular for the Haar measure $\mu$) we define a \emph{moment operator}
 \begin{gather*}
T_{\nu,t}\coloneqq\int_{\textcolor{black}{U(d)}} d\nu(U) U^{t, t},\,\,\mathrm{where}\,
U^{t,t}=U^{\otimes t}\otimes \bar{U}^{\otimes t},
\end{gather*}
\textcolor{black}{where $\bar{U}$ is entry-wise conjugation of $U$.}

One can easily show that  $0\leq\delta(\nu_{\mathcal{S}},t) \leq 1$ \cite{oszmaniec}. When $\delta(\nu_{\mathcal{S}},t)=0$ we say that $\mathcal{S}$ is {\it an exact $t$-design} and when $\delta(\nu_{\mathcal{S}},t)=1$ we say that $\mathcal{S}$ is {\it not a $t$-design}.  

\textcolor{black}{Let us note that the entries of $U^{t,t}$ are monomials of the order $t$ in the entries of $U$ and of the order $t$ in the entries of $\bar{U}$. We will call them $(t,t)$-monomials. The space of $(t,t)$-polynomials, which we denote by $\mathcal{H}_t$, is defined as the linear span of $(t,t)$-monomials. One can write any element $f_A\in\mc{H}_t$ as 
\begin{gather*}
    f_{A}(U)=\mathrm{Tr}\left(A U^{t,t}\right),
\end{gather*}
where $A$ is a $d^{2t}\times d^{2t}$ matrix.  Assuming that $\{\mathcal{S},\nu_\mc{S}\}$ is a $\delta$-approximate $t$-design we have
\begin{gather*}
\left|\int_{U(d)}d\nu_\mc{S}(U)f_A(U)-\int_{U(d)}d\mu(U) f_A(U)\right|=\left|\tr\left(A\left(T_{\nu_\mc{S},t}-T_\mu,t\right)\right)\right|\leq \|A\|_1\delta(\nu_\mc{S},t),
\end{gather*}
where $\|A\|_1=\tr\sqrt{AA^\dagger}$. Thus $\delta(\nu_\mc{S},t)$ controls the error we make when taking average of $f_A$ with respect to $\nu_\mc{S}$ instead of the Haar measure.}

A map $U\mapsto U^{t,t}$ is a representation of the unitary group $U(d)$. This representation is reducible and decomposes into some irreducible representations $U(d)$.

\begin{fact}(\cite{Stroomer}) \label{fact:ut_decomp}
Irreducible representations that appear in the decomposition of $U\mapsto U^{\otimes t}\otimes \bar{U}^{\otimes t}$ are $\pi_\lambda$ with $l(\lambda)=d$, $\Sigma(\lambda)=0$ and $\Sigma(\lambda_+) \leq t$. That is we have 
\begin{gather}\label{decomposition}
    U^{\otimes t}\otimes \bar{U}^{\otimes t}\simeq \bbone^{\oplus m_0}\oplus \bigoplus_{\lambda\in \Lambda_t}\pi_\lambda(U)^{\oplus m_{\lambda}}\simeq  \left(U\otimes \bar{U}\right)^{\otimes t},
\end{gather}
where 
\begin{gather}
    \label{eq:big-lambda}
    \Lambda_t=\left\{\lambda=(\lambda_1,\ldots,\lambda_d)|\,\lambda\in\mathbb{Z}^d,\lambda\neq 0,\,\forall_k\,\lambda_k\geq\lambda_{k+1},\,\Sigma(\lambda)=0,\,\Sigma(\lambda_{+})\leq t\right\},
\end{gather}
and $\bbone$ stands for the trivial representation and $m_0$ is its  multiplicity and $m_{\lambda}$ is the multiplicity of $\pi_\lambda$ \textcolor{black}{and $\simeq$ stands for a unitary equivalence of representations}.

\end{fact}

The representations occurring in decomposition \eqref{decomposition} are in fact irreducible representation of the projective unitary group, $PU(d)=U(d)/\sim$, where $U\sim V$ iff $U=e^{i\phi} V$. One can show that every irreducible representation of $PU(d)$ arises this way for some, possibly large, $t$ \cite{Dieck}. For $t=1$ decomposition (\ref{decomposition}) is particularly simple and reads $U\otimes \bar{U}\simeq\mathrm{Ad}_U\oplus \bbone$, where $\bbone$ stands for the trivial representation and $\mathrm{Ad}_U$ is the adjoint representation of $U(d)$ and $PU(d)\simeq\mathrm{Ad}_{U(d)}$\footnote{By $\mathrm{Ad}_U$ we mean the matrix $\mathrm{Ad}_U:\mathfrak{su}(d)\rightarrow \mathfrak{su}(d)$, $\mathrm{Ad}_U(X)=UXU^{-1}.$}.

For any irreducible representation $\pi_{\lambda}$, $\lambda \in \Lambda_t$ we define
\begin{gather*}
    T_{\nu_{\mathcal{S}},\lambda}=\int_{U(d)} d\nu_{\mathcal{S}}(U)\pi_\lambda(U).
\end{gather*}
Next we define $\delta(\nu_{\mathcal{S}},\lambda)\coloneqq  \|T_{\nu_{\mathcal{S}},\lambda}\|$. It follows directly from the  definitions and discussion above that 
\begin{gather}\label{eq:delta_t}
    T_{\nu_{\mathcal{S}},t}\simeq \bigoplus_{\lambda\in \Lambda_t}\left (T_{\nu_{\mathcal{S}},\lambda}\right)^{\oplus m_\lambda},\;
    \delta(\nu_S,t) = \mathrm{sup}_{\lambda \in \Lambda_t}\delta(\nu_{\mathcal{S}},\lambda).
\end{gather}
 One can also define $\delta(\nu_\S):=\mathrm{sup}_{t}\delta(\nu_{\mathcal{S}},t)$. It is known that \textcolor{black}{for $\mathcal{S}$ finite and $\nu_\mathcal{S}$ uniform } $\delta_\mathrm{opt} (\S)\leq\delta(\nu_\S)\leq 1$, where $\delta_\mathrm{opt} = \frac{2\sqrt{|\mc{S}| - 1}}{|\mc{S}|}$ \cite{sarnak}.

\begin{lemma}\label{lemma:partitions}
Assume $\lambda \in \Lambda_t$. Then $\|\lambda\|_1=2k$, where the integer $k$ satisfies $1\leq k\leq t$. Moreover the number of distinct irreducible representations $\pi_\lambda$ with $\|\lambda\|_1 =2k$ is given by
\begin{gather}\label{smalld}
\color{black}
    \alpha_{2k} = \left\{
        \begin{array}{ll}
            p(k)^2  & d\geq 2k,\\
		\sum_{n=1}^{k} p_n(k)\tilde{p}_{d-n}(k)& k+1 \leq d<2k,\\
            \sum_{n=1}^{d-1} p_n(k)\tilde{p}_{d-n}(k)& 2\leq d\leq k.
	\end{array}
    \right.
\end{gather}
where $p_{n}(k)$ is the number of partitions of $k$ with exactly $n$ integers and $\tilde{p}_{n}(k)$ is the number of partitions of $k$ with at most $n$ integers. When $d\geq 2t$ formula (\ref{smalld}) simplifies to 
\begin{gather}\label{samlld}
\alpha_{2k}=p(k)^2,
\end{gather}
where $p(k)$ is number of all partitions of $k$. 
\end{lemma}
\begin{proof}
First, we will prove $\|\lambda\|_1=2k$. We define $\lambda_-$ to be the subsequence of $\lambda$ of negative integers. Then from the condition $\Sigma(\lambda) = 0$ we have:
\begin{equation*}
    \Sigma(\lambda_-) = (\Sigma(\lambda_-) + \Sigma(\lambda_+)) - \Sigma(\lambda_+) = \Sigma(\lambda) - \Sigma(\lambda_+) = - \Sigma(\lambda_+).
\end{equation*}
Next, from the Eq. \ref{eq:lamb-norm}:
\begin{equation*}
    \| \lambda \|_1 = \sum_{k=1}^d| \lambda_k | = \Sigma(\lambda_+) - \Sigma(\lambda_-) = 2\Sigma(\lambda_+) \leq 2t.
\end{equation*}
Thus $\| \lambda \|_1 = 2k$ for $k = \Sigma(\lambda_+)$. Now let us put $n=l(\lambda_+)$ then the $\lambda_+$ is a decreasing sequence of $n$ positive integers that sum up to $\Sigma(\lambda_+) = k$ so it is a partition of $k$ with exactly $n$ integers and $(-(\lambda_-)_{l(\lambda_-)}, ... , -(\lambda_-)_1)$ is a decreasing sequence of $l(\lambda_-) \le d - n$ positive integers that sum up to $-\Sigma(\lambda_-) = k$ so it is a partition of $k$ with at most $n$ integers. \\
\textcolor{black}{On the other hand, if we take $\eta$ to be a partition of $k$ with exactly $n$ integers and $\zeta$ to be a partition of $k$ with at most $n$ integers such that $d \ge n + l(\zeta)$ then they can be combined into a sequence}: 
\begin{equation*}
    \tilde{\lambda} = (\eta_1, ..., \eta_n, \underbrace{0, ..., 0}_{d-n-l(\zeta)}, -\zeta_{l(\zeta)}, ..., -\zeta_1),
\end{equation*}
that is clearly an element of $\Lambda_t$ and $\| \tilde{\lambda} \|_1 = 2k$. Hence there is a one to one correspondence between sequences like $\lambda$ and pairs of partitions like $(\eta, \zeta)$. Thus to \textcolor{black}{prove} the formula \eqref{smalld} the only thing left to do is to note that the equations $\Sigma(\lambda) = 0$ and $\lambda \neq 0$ imply inequalities $n \ge 1$ and $n \le d - l(\lambda_-) \le d-1$.\\
The formula \eqref{samlld} results easily from the fact that for $n > k$ we have $p_n(k) = 0$ and $\tilde{p}_{n}(k)=p(k)$ for $n\geq k$.
\end{proof}
\begin{lemma}\label{lemma:norm_bound}
Let $U,V\in U(d)$. Assume $\lambda \in \Lambda_t$. Then 
\begin{gather*}
    \|\pi_\lambda(U)-\pi_\lambda(V)\|\leq\frac{\pi}{2}\|\lambda\|_1 \|U-V\| .
\end{gather*}
\end{lemma}
\begin{proof}
As the operator norm is \textcolor{black}{unitarily} invariant it is enough to show that for a diagonal unitary matrix $U=\mathrm{diag}\left(e^{i\phi_1},\ldots,e^{i\phi_d}\right)\in U(d)$, $\phi_k\in (-\pi,\pi]$ we have:
\begin{gather*}
    \|\pi_\lambda(U)-\pi_\lambda(\bbone)\|\leq \frac{\pi}{2} \|\lambda\|_1 \|U-\bbone\| .
\end{gather*}
One easily sees that 
\begin{gather*}
 \|U-\bbone\|=2\sup_{1\leq k\leq d}\left|\sin\left(\frac{\phi_k}{2}\right)\right|,\\
 \frac{2}{\pi} \sup_{1\leq k\leq d} |\phi_k| \leq \|U-\bbone\|
 \leq \sup_{1\leq k\leq d} |\phi_k|.
\end{gather*}
The eigenvalues of $\pi_{\lambda}(U)$ are given by 
\begin{gather*}
   e^{i\psi_k},\,\, \psi_k=\sum_{i=1}^d \lambda_{i,k}\phi_i,
\end{gather*}
where $\lambda_k=(\lambda_{1,k},\ldots,\lambda_{d,k})$ satisfies $\|\lambda_k\|_1\leq\|\lambda\|_1$ (see Lemma \ref{weight}). Thus 
\begin{gather*}
    \|\pi_\lambda(U)-\bbone\|=2\sup_{1\leq k\leq d_\lambda}\left|\sin\left(\frac{\psi_k}{2}\right)\right|\leq \sup_{1\leq k\leq d_\lambda} |\psi_k|\leq\|\lambda\|_1\sup_{1\leq k\leq d} |\phi_k|\leq \frac{\pi}{2}\|\lambda\|_1\|U-\bbone\|.
\end{gather*}

\end{proof}

\color{black}

\subsection{Moment operators for quantum circuits}
We can generalize the notion of moment operators to random quantum circuit in a natural way. Recall that a quantum circuit of depth $m$ is a product of $m$ quantum gates. Thus we can identify quantum circuits with elements of $U(d)^{\times m}$.

Consider an ensemble $\{ \mathcal{R}, \nu_\mathcal{R} \}$ where $\mathcal{R} \subset U(d)^{\times m}$ is a set of circuits of length $m$ and $\nu_\mathcal{R}$ is a probability measure on $\mathcal{R}$. For $\mathbf{U} = (U_1, ..., U_m) \in U(d)^{\times m}$ let us define 
\begin{equation}
\label{eq:poly_circuit}
    \mathbf{U}^{t, t} := U_1^{t, t} \otimes ... \otimes U_m^{t, t}.
\end{equation}
Note that the entries of $\mathbf{U}^{t, t}$ form a basis of the space of all $(t,t)-$polynomials in entries of $U_1, ... , U_m$ which we will call $\mathcal{H}_{t, m}$. The average of $\mathbf{U}^{t, t}$ over all circuits with the same probability is
\begin{gather*}
    T_{\mu^{\times m}, t} := \int_{U(d)^m}d\mu^{\times m}(\mathbf{U}) \mathbf{U}^{t, t} = \int_{U(d)} d\mu(U_1) ... \int_{U(d)} d\mu(U_m) \mathbf{U}^{t, t} =\\
    = \left( \int_{U(d)}d\mu(U_1) U_1^{t, t} \right) \otimes ... \otimes \left( \int_{U(d)}d\mu(U_m) U_m^{t, t} \right) = T_{\mu, t}^{\otimes m}.
\end{gather*}
where $\times$ is a product of measures. Using Fact \ref{fact:ut_decomp} we get that:
\begin{equation*}
    \mathbf{U}^{t, t} \simeq \bigotimes_{i=1}^m \left( \bbone^{\oplus m_0}\oplus \bigoplus_{\lambda\in \Lambda_t}\pi_\lambda(U_i)^{\oplus m_{\lambda}} \right) \simeq \bigoplus_{\lambda_1, ..., \lambda_m \in \tilde{\Lambda}_t} \bigotimes_{i=1}^m \pi_{\lambda_i}(U_i)^{\oplus m_{\lambda_i}},
\end{equation*}
where $\tilde\Lambda_t := \Lambda_t \cup \{0\}$. To simplify notation let us define:
\begin{gather*}
    \bm{\lambda} := (\lambda_1, ..., \lambda_m) \in \tilde\Lambda_t^{\times m}, \\
    \Lambda_{t, m} := \tilde\Lambda^{\times m}_t \setminus \{ (0, ..., 0) \}, \\
    \pi_{\bm{\lambda}} (\mathbf{U}) := \pi_{\lambda_1}(U_1) \otimes ... \otimes \pi_{\lambda_m}(U_m),\\
    T_{\nu_\mathcal{R}, t} := \int_{U(d)^m} d\nu_\mathcal{R}(\mathbf{U}) \mathbf{U}^{t, t},\\
    T_{\nu_\mathcal{R}, \bm{\lambda}} := \int_{U(d)^m} d\nu_\mathcal{R}(\mathbf{U}) \pi_{\bm{\lambda}} (\mathbf{U}), \\
    \delta(\nu_\mathcal{R}, \bm{\lambda}) := \| T_{\nu_\mathcal{R}, \bm{\lambda}} \|.
\end{gather*}
It is easy to see that:
\begin{equation*}
    \delta(\nu_\mathcal{R}, t) := \| T_{\mu, t}^{\otimes m} - T_{\nu_\mathcal{R}, t} \| = \sup_{\bm{\lambda} \in \Lambda_{t, m}} \| T_{\nu_\mathcal{R}, \bm{\lambda}} \| = \sup_{\bm{\lambda} \in \Lambda_{t, m}} \delta(\nu_\mathcal{R}, \bm{\lambda}).
\end{equation*}

\color{black}

\color{black}
\section{Frobenius-Schur indicators}
\label{sec:fs-indicators}

In Section \ref{sec:inequalities_review} we explained that to obtain bounds on the norm of the random matrix $X = \sum_{k} X_k$ one needs to compute all moments $\mathbb{E}\left( X_k^n \right)$. In this article we will be interested in estimating $\delta(\nu_\mathcal{S}, t_1)$ for $\nu_\mathcal{S}$ uniform and $\mathcal{S}$ finite with each element of $\mathcal{S}$ chosen from an exact $t_2$-design $\{ \mathcal{D}, \nu_\mathcal{D} \}$. Thus in our scenario the role of $X$ is played by
\begin{equation*}
    T_{\nu_\mathcal{S}, \lambda_{}} = \int_{U(d)} d\nu_\mathcal{S}(U) \pi_\lambda(U) = \frac{1}{|\mathcal{S}|} \sum_{U \in \mathcal{S}} \pi_{\lambda}(U),
\end{equation*}
$X_k$'s are matrices proportional to $\pi_\lambda(U)$ and the average is taken over $\nu_\mathcal{D}$
\begin{equation*}
    \E{\nu_\mathcal{D}}\pi_\lambda(U^n) := \int_{U(d)} d\nu_\mathcal{D} (U) \pi_\lambda(U)^n,
\end{equation*}
for $\lambda \in \Lambda_{t_1}$ and $n \in \mathbb{Z}$. In particular in Lemma \ref{un} we show that this integral is proportional to $\bbone_{d_\lambda}$ and the proportionality constant is the Frobenius-Schur indicator $\int_{U(d)} d\mu(U) \chi_\lambda(U^n)$ divided by $d_\lambda$.

\begin{lemma}
\label{lemma:td_eq_haar}
    Consider an irreducible finite-dimensional representation $\pi_\lambda$ of $U(d)$ with the highest weight $\lambda \in \Lambda_t$, an integer $n$ and an exact $(|n|\cdot t)$-design $\{ \mathcal{D}, \nu_\mathcal{D}\}$. Then the average of $\pi_\lambda(U)^n$ taken over $\nu_\mathcal{D}$ and $\mu$ is the same, that is
    \begin{equation*}
       \E{\nu_\mathcal{D}}\pi_\lambda(U^n) = \int_{U(d)} d\nu_\mathcal{D} (U) \pi_\lambda(U)^n = \int_{U(d)} d\mu (U) \pi_\lambda(U)^n = \E{\mu}\pi_\lambda(U^n)
    \end{equation*}
\end{lemma}
\begin{proof}
     If $n < 0$ we can use unitarity of $\pi_\lambda$ to obtain that for any measure $\nu$
    \begin{equation*}
        \int_{U(d)} d\nu (U) \pi_\lambda(U)^{n} = \int_{U(d)} d\nu (U) \pi_\lambda(U)^{-|n|} = \left( \int_{U(d)} d\nu(U) \pi_\lambda(U)^{|n|} \right)^\dagger.
    \end{equation*}
    Therefore it is enough to prove the Lemma for $n \ge 0$. In such case note that
    \begin{equation*}
        U^{\otimes nt} \otimes \bar{U}^{\otimes nt} \simeq \left[ U^{\otimes t} \otimes \bar{U}^{\otimes t} \right]^{\otimes n} \simeq \left[ \bigoplus_{\lambda\in \Lambda_t}\pi_\lambda(U)^{\oplus m^{\lambda}} \right]^{\otimes n}.
    \end{equation*}
    Entries of the right-hand side operator above are
    \begin{equation}
    \label{eq:nt_entries}
        \pi_{\lambda_1}(U)_{i_1 j_1} \cdot \pi_{\lambda_2}(U)_{i_2 j_2} \cdot ... \cdot\pi_{\lambda_n}(U)_{i_n j_n},
    \end{equation}
    for all possible $\lambda_1, ..., \lambda_n \in \Lambda_t$ and $1 \le i_k, j_k \le d_{\lambda_k}$. Since $\nu_\mathcal{D}$ is a $(n \cdot t)$-design integrating the expression \eqref{eq:nt_entries} over $\nu_\mathcal{D}$ and $\mu$ gives the same result. Thus the Lemma follows from the fact that the entries of $\pi_\lambda(U)^n$ are linear combinations of expressions \eqref{eq:nt_entries} with $\lambda_1 = ... = \lambda_n = \lambda$ and $j_k = i_{k+1}$.
\end{proof}

\begin{lemma}\label{un}
For any irreducible representation $\pi_{\lambda}$ of a compact Lie group $G$ and $n \in \mathbb{Z}$ we have 
\begin{gather*}
    \int_{G}\pi_{\lambda}(U^n)d\mu(U)=\delta_\lambda(n) \bbone_{d_\lambda},\\\nonumber
    \delta_\lambda(n) := \frac{1}{d_\lambda}\int_{G}\chi_\lambda(U^n)d\mu(U).
\end{gather*}
\end{lemma}
\begin{proof}
Assume that $n \ge 0$. For any $g\in G$ we have
\begin{gather*}
    \pi_{\lambda}(g)\int_{G}\pi_{\lambda}(U^n)d\mu(U)=\int_{G}\pi_{\lambda}(gU^n)d\mu(U)=\int_{G}\pi_{\lambda}(gVgV\ldots gVg)d\mu(V),
\end{gather*}
where in the last equality we performed a change of variables $U=Vg$ and used the invariance of the Haar measure. Similarly 
\begin{gather*}
    \left(\int_{G}\pi_{\lambda}(U^n)d\mu(U)\right)\pi_\lambda(g)=\int_{G}\pi_{\lambda}(U^ng)d\mu(U)=\int_{G}\pi_{\lambda}(gVgV\ldots gVg)d\mu(V),
\end{gather*}
where in the last equality we performed a change of variables $U=gV$ and used the invariance of the Haar measure. In case $n < 0$ the argument is analogous but with the change of variables $U = g^{-1}V$ in the first equation and $U = Vg^{-1}$ in the second.
Thus for any $g\in G$, the matrix $\pi^\lambda(g)$ commutes with the integral in (\ref{un}). By Schur lemma this integral must be proportional to identity. The proportionality constant, $\delta_\lambda(n)$, can be establish by taking trace of both sides of equation (\ref{un}).
\end{proof}

\begin{corollary} \label{corollary:td_eq_del}
    For any irreducible finite-dimensional representation $\pi_\lambda$ of $U(d)$ with the highest weight $\lambda \in \Lambda_t$, an integer $n$ and an exact $(|n|\cdot t)$-design $\{ \mathcal{D}, \nu_\mathcal{D}\}$ we have
    \begin{equation*}
        \E{\nu_\mathcal{D}}\pi_\lambda(U^n) = \delta_\lambda(n) \bbone_{d_\lambda}.
    \end{equation*}
\end{corollary}
\begin{proof}
    From Lemma \ref{lemma:td_eq_haar} we know that
    \begin{equation*}
        \E{\nu_\mathcal{D}}\pi_\lambda(U^n) = \E{\mu}\pi_\lambda(U^n),
    \end{equation*}
    and from Lemma \ref{un} that
    \begin{equation*}
        \E{\mu}\pi_\lambda(U^n) = \delta_\lambda(n) \bbone_{d_\lambda}.
    \end{equation*}
\end{proof}

From the orthogonality of characters we know that for $\lambda \neq 0$
\begin{equation*}
    \int_G d\mu(U) \chi_\lambda(U) = \int_G d\mu(U)\, 1 \cdot \chi_\lambda(U) = \int_G d\mu(U) \overline{\chi_0(U)}  \chi_\lambda(U) = 0,
\end{equation*}
thus $\delta_\lambda(\pm 1) = 0$. In case of $\delta_\lambda(\pm 2)$ we use a well known fact that 
\begin{equation*}
    \int_{G}\chi_\lambda \left( U^{\pm 2} \right)d\mu(U)
\end{equation*}
is equal to $1$, $0$ or $-1$ for $\pi_\lambda$ real, complex or quaternionic respectively. Thus
\begin{equation*}
    \delta_\lambda(\pm 2) = \begin{cases}
        \frac{1}{d_\lambda} &\mathrm{if} \, \pi_\lambda \, \mathrm{real}\\
        0 &\mathrm{if} \, \pi_\lambda \, \mathrm{complex}\\
        -\frac{1}{d_\lambda} &\mathrm{if} \, \pi_\lambda \, \mathrm{quaternionic}\\
    \end{cases}.
\end{equation*}
To calculate $\delta_\lambda(n)$ for $|n| > 2$ we will use the following result from \cite{abuhamed2007}.

\begin{fact}
\label{fact:fs_indicator}
Let $G$ be a finite dimensional semisimple Lie group. Let $\pi_{\lambda^s}$ be an irreducible representation of $G$ with highest weight $\lambda^s$, $\mathcal{W}$ the Weyl group of $\mathfrak{g}$ and $\rho$ the half sum of positive roots. Then for $n \neq 0$
\begin{equation*}
    \label{eq:abuhamed_formula}
    \delta_{\lambda^s}(n) = \frac{1}{d_{\lambda^s}} \sum_{\sigma \in \mathcal{W}} \mathrm{sgn}(\sigma) \; m_{\lambda^s}\left(\frac{\rho - \sigma \cdot \rho}{n}\right).
\end{equation*}
\end{fact}
Immediate conclusion from Fact \ref{fact:fs_indicator} is that there is $n_0$ such that for $n\geq n_0$ elements $\frac{\rho - \sigma \cdot \rho}{n}$ are integral elements if and only if $\sigma \cdot \rho = \rho$. In the next lemma we calculate $n_0$ for the group $SU(d)$.
\begin{lemma}
\label{lemma:abuhamed_simplified}
For the group $SU(d)$ the constant $n_0$ is equal to $d+1$ and for $n \ge n_0$ the formula \eqref{eq:abuhamed_formula} simplifies to:
\begin{equation}
    \label{eq:abuhamed_simplified}
    \delta_{\lambda^s}(n) = \frac{m_{\lambda^s}(0)}{d_{\lambda^s}}.
\end{equation}
\end{lemma}
\begin{proof}
First, let us calculate $\rho$:
\begin{equation*}
    \rho = \frac{1}{2} \sum_{\alpha \in \Phi^+} \alpha = 
    \frac{1}{2} \sum_{1 \le i < j \le d} \alpha_{i, j} = 
    \frac{1}{2} \sum_{1 \le i < j \le d} L_i - L_j =
    \sum_{i=1}^d \left( \frac{d-1}{2}-i+1 \right)L_i.
\end{equation*}
Thus for $i \neq j$ we have $\rho_i \neq \rho_j$ and the only $\sigma \in S_d$ such that $\sigma \cdot \rho = \rho$ is the identity, which proves the simplified formula \eqref{eq:abuhamed_simplified}. In order to find minimal $n_0$ note first that
\begin{equation*}
    (\rho - \sigma \cdot \rho)_i = \rho_i - \rho_{\sigma(i)} = \sigma(i) - i,
\end{equation*}
and that for $\frac{\rho - \sigma \cdot \rho}{n}$ to be an integral element it is required that for all $1 \le i \le d-1$
\begin{equation*}
    \braket{\rho - \sigma \cdot \rho | \alpha_{i, i+1}} = 
    (\rho - \sigma \cdot \rho)_i - (\rho - \sigma \cdot \rho)_{i+1} =
    \sigma(i) - \sigma(i+1) + 1,
\end{equation*}
is divisible by $n$ or equivalently
\begin{equation}
    \label{eq:c_const}
    \forall_{1 \le i \le d-1} \exists_{C_i \in \mathbb{Z}}\quad \sigma(i+1) - \sigma(i) = C_i n + 1.
\end{equation}
If $n=d$ then $\sigma(i) = i - 1 \;\mathrm{mod}\; d$ satisfies this condition thus $n_0 > d$. On the other hand, for $n > d$ the condition $|\sigma(i+1)-\sigma(i)| \le d-1$ implies that the constant $C_i$ in \eqref{eq:c_const} has to be zero for all $i$. Let us choose $j$ such that $j \neq d$ and $\sigma(j)=d$ then
\begin{equation*}
    1 = C_{j} n + 1 = \sigma(j+1) - \sigma(j) = \sigma(j+1)-d \quad\Rightarrow\quad \sigma(j+1) = d+1,
\end{equation*}
what is an obvious contradiction and $\sigma^{-1}(d)$ has to be $d$ but this implies $\sigma(d-1) = \sigma(d)-1=d-1$ and analogously for all $1 \le i \le d$ we have $\sigma(i)=i$. Therefore for $n > d$ the only $\sigma$ satisfying \eqref{eq:c_const} is the trivial one.
\end{proof}
In the next lemma we show when the representation $\pi_{\lambda^s}$ has the weight $0$. In particular it implies that for every $\lambda \in \Lambda_t$ the representation $\pi_{\lambda^s}$ has the weight $0$.
\begin{lemma}\label{lemma:projective}
Suppose $\pi_{\lambda^s}$ is a representation of $SU(d)$ with the highest weight $\lambda^s$. Then the following are equivalent:
\begin{enumerate}
    \item $\pi_{\lambda^s}$ has weight $0$,
    \item $\frac{1}{d}\sum_{j=1}^{d-1}j\lambda^s_{j}$ is an integer,
    \item there exists an irreducible finite-dimensional representation of $U(d)$ with highest weight $\lambda$ - $\pi_\lambda$ such that $\Sigma(\lambda)=0$ and $\pi_\lambda|_{SU(d)}=\pi_{\lambda^s}$.
\end{enumerate}
\end{lemma}
\begin{proof}
We will start with 1) $\Rightarrow$ 2).\\
From Fact \ref{fact:possible_weights} there exist $c_1, ..., c_{d-1} \in \mathbb{Z}$ such that
\begin{equation*}
    \lambda^s = \lambda^s - 0 = \sum_{i=1}^{d-1}c_i\alpha_{i, i+1},
\end{equation*}
thus
\begin{equation*}
    \lambda^s_j = \braket{\lambda^s|\alpha_{j, j+1}} = \braket{\sum_{i=1}^{d-1}c_i\alpha_{i, i+1} | \alpha_{j, j+1}} =
    \sum_{i=1}^{d-1}c_i\braket{\alpha_{i, i+1}|\alpha_{j, j+1}} =
    -c_{j-1} + 2 c_j - c_{j+1},
\end{equation*}
where we assume $c_0 = 0 = c_{d}$. It follows
\begin{gather*}
    \frac{1}{d}\sum_{j=1}^{d-1}j\lambda^s_{j} =
    \frac{1}{d}\sum_{j=1}^{d-1}j(-c_{j-1} + 2 c_j - c_{j+1}) =\\=
    \frac{1}{d} \left\{ \sum_{j=1}^{d-2} \left[ -(j-1) + 2j - (j+1) \right] c_j + \left[ -(d-2) + 2(d-1) \right] c_{d-1} \right\} =\\=
    \frac{d}{d}c_{d-1} = c_{d-1} \in \mathbb{Z}.
\end{gather*}

2) $\Rightarrow$ 3)\\
For any $m \in \mathbb{Z}$
\begin{equation*}
    \lambda = (m + \sum_{i=1}^{d-1} \lambda^s_{i},\; m + \sum_{i=2}^{d-1}\lambda^s_{i},\; ...\;,\; m + \lambda^s_{d-1},\; m),
\end{equation*}
satisfies $\lambda_i \in \mathbb{Z}$ and $i < j \,\Rightarrow\, \lambda_i \ge \lambda_j$. Therefore there exists an irreducible finite-dimensional representation of $U(d)$ with highest weight $\lambda$ - $\pi_\lambda$. From Lemma \ref{lemma:young_diagrams} clearly $\pi_\lambda|_{SU(d)} = \pi_{\lambda^s}$ and for $m = -\frac{1}{d}\sum_{j=1}^{d-1}j\lambda^s_{j}$ we have $\Sigma(\lambda) = 0$.\\

3) $\Rightarrow$ 1)\\
Let us choose $\forall_{\sigma \in S_d}\; t_\sigma = \frac{1}{d!}$ then $\sum_{\sigma \in S_d}t_\sigma = 1$ and
\begin{gather*}
    \sum_{\sigma \in S_d} t_\sigma \sigma \cdot \lambda = 
    \frac{1}{d!}(\sum_{\sigma \in S_d}\lambda_{\sigma(1)},\; ... \;,\; \sum_{\sigma \in S_d}\lambda_{\sigma(d-1)}) = \frac{1}{d}(\Sigma(\lambda),\; ... \;,\; \Sigma(\lambda)) = 0.
\end{gather*}
What is more for $c_1 = \lambda_1$ and $\forall_{2 \le i \le d-1}\; c_i = \lambda_{i-1}+\lambda_i$ we have
\begin{equation*}
    \sum_{i=1}^{d-1}c_i \alpha_{i, i+1} = \lambda = \lambda - 0.
\end{equation*}
Hence by Fact \ref{fact:possible_weights} weight $0$ is in $\pi_\lambda$ what implies that weight $0$ is in $\pi_{\lambda^s}$.

\end{proof}

As an example we next explicitly calculate the value of $\delta_\lambda(n)$ for $G=SU(2)$.

\begin{lemma}\label{lambda}
The constant in Lemma \ref{un} for $SU(2)$ and non-trivial $\pi_\lambda$ is given by
\begin{gather*}
    \delta_\lambda(n)=\begin{cases}
        1 & \text{if $n=0$,}\\
        0 & \text{if $n=\pm 1$,}\\
        \frac{1}{d_\lambda} & \text{if $n\in\mathbb{Z}\smallsetminus\{-1,0,1\}$ and $\lambda^s$-even, }\\
        -\frac{1}{d_\lambda} & \text{if $n=\pm 2$ and $\lambda^s$-odd}\\
        0 & \text{if $n\in\mathbb{Z}\smallsetminus\{-2,0,2\}$ and $\lambda^s$-odd}
    \end{cases} 
\end{gather*}
\end{lemma}
\begin{proof}
The result for $n=0$ and $n=1$ follows from definition of $\delta_\lambda$ and irreducibility of $\pi_{\lambda}$. For remaining $n$'s we will use Fact \ref{fact:fs_indicator}.
First, let us notify that for $SU(2)$:
\begin{equation*}
    \Delta = \{\alpha_{12}\}, \qquad \alpha_{12} = 2, \qquad \rho = 1, \qquad [1\;2] \cdot \mu = -\mu,
\end{equation*}
then from Kostant formula \ref{fact:kostant}:
\begin{equation}
\label{eq:multiplicity}
    m_{\lambda^s}(\mu) = \sum_{\sigma \in S_2} \sgn{\sigma}\, p(\sigma \cdot (\lambda^s+\rho) - (\mu+\rho)) = p(\lambda^s - \mu) - p(-\lambda^s - \mu - 2).
\end{equation}
Recall that $p(m)$ is the number of ways $m$ can be expressed as a linear combinations of positive simple roots with non-negative integer coefficients so in our case
\begin{equation*}
    p(m) = \begin{cases}
        1 & \text{if $m$-even and non-negative,}\\
        0 & \text{if $m$-odd or negative.}
    \end{cases} 
\end{equation*}
This combined with the fact that $-\lambda^s - \mu - 2 > 0$ implies $\| \mu \|_1 > \| \lambda \|_1$ and Lemma \ref{weight} simplifies \eqref{eq:multiplicity} to
\begin{equation*}
    m_{\lambda^s}(\mu) = 
    p(\lambda^s - \mu) = \begin{cases}
        1 & \text{if $(\lambda^s - \mu)$-even and non-negative,}\\
        0 & \text{if $(\lambda^s - \mu)$-odd or negative.}
    \end{cases} 
\end{equation*}
Now, we are ready to use formula (\ref{eq:abuhamed_formula}) which for $SU(2)$ has a form
\begin{equation*}
    \delta_{\lambda^s}(n) = \frac{1}{d_{\lambda^s}} \left( m_{\lambda^s}(0) - m_{\lambda^s}\left(\frac{2}{n} \right) \right).
\end{equation*}
Hence
\begin{equation*}
    \delta_{\lambda^s}(\pm2) = \frac{1}{d_{\lambda^s}} \left( m_{\lambda^s}(0) - m_{\lambda^s}(\pm1 ) \right) =
    \begin{cases}
        \frac{1}{d_{\lambda^s}} & \text{if $\lambda^s$-even,}\\
        -\frac{1}{d_{\lambda^s}} & \text{if $\lambda^s$-odd.}
    \end{cases}
    ,
\end{equation*}
and for $|n| \ge 3$
\begin{equation*}
    \delta_{\lambda^s}(n) = \frac{1}{d_{\lambda^s}} m_{\lambda^s}(0) =
    \begin{cases}
        \frac{1}{d_{\lambda^s}} & \text{if $\lambda^s$-even,}\\
        0 & \text{if $\lambda^s$-odd.}
    \end{cases}
\end{equation*}

\end{proof}

\color{black}
\section{Main results}\label{sec:main_results}
In this section we will consider two types of gate-sets:
\color{black}
\begin{enumerate}
    \item $\mathcal{S}=\{U_1,\ldots,U_n\}$, where $U_k$'s are independent random unitaries from an exact $t$-design $\mathcal{D}$ chosen according to the measure $\nu_\mathcal{D}$. Such $\mc{S}$ will be called {\it $t$-random  gate-set}.
    \item $\mathcal{S}=\{U_1,\ldots,U_n\}\cup\{U_1^{-1},\ldots,U_n^{-1}\}$, where $U_k$'s are independent random unitaries from an exact $t$-design $\mathcal{D}$ chosen according to the measure $\nu_\mathcal{D}$. Such $\mc{S}$ will be called {\it symmetric $t$-random  gate-set}.
\end{enumerate}
Note that in the symmetric case we have $U_k \in \mathcal{D}$ but not necessarily $U_k^{-1} \in \mathcal{D}$ so to obtain a symmetric $t$-random gate-set of size $2n$ we draw $n$ gates from $\{ \mathcal{D}, \nu_\mathcal{D} \}$ and then we add to the set their inverses. We also want to point out that since $\{ U(d), \mu \}$ is an exact $t$-design for any $t$ all theorems we present below apply to gate-sets where gates are Haar random or Haar random with inverses (we could call them $\infty$-random gate-sets or symmetric $\infty$-random gate-sets).\\ 
\color{black}
In order to simplify the notation we often denote the cardinality of $\mc{S}$ by $\mathcal{S}$ (instead of $|\mc{S}|$). The moment operators associated with the above two types of random sets of gates are random matrices. When $\mc{S}$ is symmetric they are actually random Hermitian matrices. Using inequalities listed in Section \ref{sec:inequalities_review} we derive upper bounds on $\mathbb{P}\left(\delta(\nu_\mc{S},\lambda)\geq \delta\right)$ and  $\mathbb{P}\left(\delta(\nu_\mc{S},t)\geq \delta\right)$. Before we proceed with concrete inequalities we note that 

\begin{lemma}
Assume that $\mathcal{S}\subset U(d)$ is a random set of quantum gates and that for every $\lambda \in \Lambda_t$ we have  $\mathbb{P}\left(\delta(\nu_\mc{S},\lambda)\geq \delta\right)\leq F(\delta,\lambda,\mc{S})$. Then 
\begin{gather*}
    \mathbb{P}\left(\delta(\nu_\mc{S},t)\geq \delta\right)\leq \sum_{\lambda \in \Lambda_t}F(\delta,\lambda,\mc{S}).
\end{gather*}
\end{lemma}
\begin{proof}
By \eqref{eq:delta_t} and using the union bound for probabilities we have
\begin{gather*}
    \mathbb{P}\left(\delta(\nu_\mc{S},t)\geq \delta\right)=\mathbb{P}\left(\mathrm{sup}_{\lambda \in \Lambda_t} \delta(\nu_{\mathcal{S}}, \lambda) \geq \delta\right) \leq \sum_{\lambda \in \Lambda_t} \mathbb{P} \left(\delta(\nu_\mc{S}, \lambda) \geq \delta\right) \leq \sum_{\lambda \in \Lambda_t}F(\delta, \lambda, \mc{S}).
\end{gather*}
\end{proof}
\noindent Thus, to find a bound on $ \mathbb{P}\left(\delta(\nu_\mc{S},t)\geq \delta\right)$ it is enough to find bounds on $\mathbb{P}\left(\delta(\nu_\mc{S},\lambda)\geq \delta\right)$, $\lambda \in \Lambda_t$, which we do in next sections.
\subsection{Bernstein type bounds}
In this section we derive bounds on $\mathbb{P}\left(\|T_{\nu_{\mc{S}},\lambda}\|\geq \delta\right)$ using Bernstein inequality (see Section \ref{sec:inequalities_review}). It is worth to mention that computationally this is the simplest derivation presented in this paper as it requires only the knowledge of the second moments. 
\begin{theorem}
\label{th:bernstein_bound}
Let \textcolor{black}{$ \lambda $ be an element of $\Lambda_t$ and} $\mathcal{S}$ be a \textcolor{black}{$t$-}random gate-set \textcolor{black}{or a symmetric $(2t)$-random gate-set and $\nu_\mc{S}$ a uniform measure}. Then
\begin{gather*}
    \mathbb{P}\left(\delta(\nu_{\mc{S}},\lambda)\geq \delta\right)\leq\begin{cases} 
   B_1(\delta,\lambda,\mc{S}),&\,\mathrm{when}\,\mathcal{S}-\textrm{symmetric \textcolor{black}{$(2t)$-}random  gate-set}\\
   B_2(\delta,\lambda,\mc{S}),&\,\mathrm{when}\,\mathcal{S}-\textrm{\textcolor{black}{$t$-}random gate-set}
   \end{cases}
\end{gather*}
    where 
\begin{gather*}
    B_1(\delta,\lambda,\mc{S})= 2d_\lambda \exp\left( {\frac{-3\mathcal{S}\delta^2}{6\left(1 + \delta_\lambda(2) \right)+4\delta}}\right),\\  
    B_2(\delta,\lambda,\mc{S})= 2d_\lambda \exp\left( {\frac{-3\mathcal{S}\delta^2}{6+2\delta}}\right),\\
    \delta_\lambda(2) = \begin{cases}
        \frac{1}{d_\lambda} &\mathrm{if} \, \pi_\lambda \, \mathrm{real}\\
        0 &\mathrm{if} \, \pi_\lambda \, \mathrm{complex}\\
        -\frac{1}{d_\lambda} &\mathrm{if} \, \pi_\lambda \, \mathrm{quaternionic}\\
    \end{cases}.
\end{gather*}    

\end{theorem}
\begin{proof}
In this proof we will use Fact \ref{fact:bernstein} where we take $X = \sum_k X_k$ to be equal $T_{\nu_\mc{S}, \lambda} = \frac{1}{\mc{S}} \sum_{U \in \mc{S}} \pi_\lambda(U)$.\\
Let us start with the Haar random case. In this case all elements of $\mc{S}$ are independent. Thus we will put $X_k = \frac{1}{\mc{S}}\pi_\lambda(U_k)$ for $k=1,...,\mc{S}$. Then \color{black} from Corollary \ref{corollary:td_eq_del}
\begin{equation*}
    \E{\nu_\mathcal{D}}X_k = \E{\nu_\mathcal{D}}\left( \frac{1}{\mathcal{S}} \pi_\lambda(U) \right) = \frac{\delta_\lambda(1)}{\mathcal{S}} \bbone_{d_\lambda} = 0.
\end{equation*}
\color{black}Since $\pi_\lambda$ is unitary we have
\begin{gather*}
    \| X_k \| = \frac{1}{\mc{S}} \| \pi_\lambda(U_k) \| = \frac{1}{\mc{S}},\\
    X_kX_k^\dagger = X_k^\dagger X_k = \frac{1}{\mc{S}^2} \bbone_{\d_\lambda}.
\end{gather*}
Hence $L=\frac{1}{\mc{S}}$ and $\nu = \| \sum_{k=1}^{\mc{S}} \E{\nu_\mathcal{D}}(X_kX_k^\dagger) \| = \frac{1}{\mc{S}}$ satisfy conditions from Fact \ref{fact:bernstein} and the result follows.

When $\mc{S}$ is symmetric we put $X_k = \frac{1}{\mc{S}}\left(\pi_\lambda\left(U\right) + \pi_\lambda \left(U^{-1}\right) \right)$\textcolor{black}{. From Corollary \ref{corollary:td_eq_del} we have $\E{\nu_\mathcal{D}}X_k = 0$.} Then using triangle inequality we get
\begin{equation*}
    \| X_k \| = \frac{1}{\mc{S}} \| \pi_\lambda(U) + \pi_\lambda(U^{-1}) \| \le \frac{2}{\mc{S}} \| \pi_\lambda(U) \| = \frac{2}{\mc{S}},
\end{equation*}
and from hermiticity of $X_k$
\begin{equation*}
    X_kX_k^\dagger = X_k^\dagger X_k = \frac{1}{\mc{S}^2} \left( \pi_\lambda\left(U^{2}\right) + 2 \bbone_{d_\lambda} + \pi_\lambda\left(U^{-2}\right) \right).
\end{equation*}
\color{black}Then using Corollary \ref{corollary:td_eq_del} once more we obtain
\begin{equation*}
    \E{\nu_\mathcal{D}} \left( X_kX_k^\dagger \right) = \frac{2}{\mathcal{S}^2} \left( \delta_\lambda(2) + 1 \right)\bbone_{d_\lambda}.
\end{equation*}
\color{black}Hence for $L = \frac{2}{\mc{S}}$ and
\begin{equation*}
    \nu = \| \sum_{k=1}^{\mathcal{S}/2} \textcolor{black}{\E{\nu_\mathcal{D}}}(X_k X_k^\dagger) \| = \frac{\delta_\lambda(2)+1}{\mc{S}},
\end{equation*}
the result follows from Fact \ref{fact:bernstein}.

\end{proof}
\subsection{Master bounds}
In this section we derive {\color{black} Theorems \ref{th:master_bound-main1} and \ref{th:master_bound-main2} from Introduction} using the formula for the master bound (see Section \ref{sec:inequalities_review}). This derivation requires knowledge of all moments and is much more accurate than the Bernstein bound given in the previous section.
\subsubsection{Haar random  gate-sets}
For \textcolor{black}{$t$-}random  gate-sets the derivation of the bound for $\mathbb{P}\left(\delta(\nu_{\mc{S}},\lambda) \geq \delta\right)$ turns out to be relatively simple.
\begin{theorem}
\label{th:master_bound}
Let \textcolor{black}{$\lambda$ be an element of $\Lambda_t$ and} $\mathcal{S}$ be a \textcolor{black}{$t$-}random gate-set \textcolor{black}{and $\nu_\mc{S}$ a uniform measure}. Then for any $\delta < 1$
\begin{gather*}
    \mathbb{P}\left(\delta(\nu_{\mc{S}},\lambda) \geq \delta\right)\leq F(\delta,\lambda,\mc{S}),
    \end{gather*}
 where    
\begin{gather*}
    F(\delta,\lambda,\mc{S})= \frac{2d_\lambda}{\left(1+\delta\right)^{\frac{\mc S}{2}\left(1+\delta\right)}\left(1-\delta\right)^{\frac{\mc S}{2}\left(1-\delta\right)}}=\frac{2d_\lambda}{(1-\delta^2)^{\frac{S}{2}}}e^{-\delta\mc{S}\,\mathrm{arctanh}(\delta)}.
\end{gather*}
\end{theorem}

\begin{proof}
To compute the master bound for non-Hermitian matrix (Fact \ref{fact:master-nonhermitian}) we need to compute $\textcolor{black}{\E{\nu_\mathcal{D}}}\exp\left(\frac{\theta}{\mathcal{S}} \mathcal{H}(\pi_\lambda(U))\right)$. First, let us note that for any unitary matrix $U \in U(d_\lambda)$:
\begin{equation*}
    \mathcal{H}(U)^2 = \begin{pmatrix} 0 & U\\ U^\dagger & 0\end{pmatrix}^2 =
    \begin{pmatrix} UU^\dagger & 0 \\ 0 & U^\dagger U \end{pmatrix} = \bbone_{2d_\lambda}.
\end{equation*}
Since $\pi_\lambda$ is unitary we have
\begin{equation*}
    \textcolor{black}{\E{\nu_\mathcal{D}}} \exp \left( \frac{\theta}{\mathcal{S}} \mathcal{H}( \pi_\lambda(U) ) \right) = \cosh{\left( \frac{\theta}{\mathcal{S}} \right)}\bbone_{2d_\lambda} + \sinh{\left( \frac{\theta}{\mathcal{S}} \right)} \textcolor{black}{\E{\nu_\mathcal{D}}} \mathcal{H}(\pi_\lambda(U)).
\end{equation*}
From \textcolor{black}{Corollary \ref{corollary:td_eq_del}} and the orthogonality of characters:
\begin{equation*}
    \textcolor{black}{\E{\nu_\mathcal{D}}}\mathcal{H}(\pi_\lambda(U)) = \mathcal{H}\left(\textcolor{black}{\E{\nu_\mathcal{D}}}\pi_\lambda(U)\right) = 0.
\end{equation*}
Hence the right hand side of the master bound is:
\begin{equation*}
    \inf_{\theta>0}e^{-\theta\delta}\mathrm{tr}\exp\left(\mathcal{S}\log \textcolor{black}{\E{\nu_\mathcal{D}}} e^{\frac{\theta}{\mathcal{S}}\mathcal{H}(X_k)}\right) = 
    \inf_{\theta > 0} 2d_\lambda \; e^{-\theta\delta} \cosh^\mathcal{S} \left( \frac{\theta}{\mathcal{S}} \right).
\end{equation*}
By calculating derivative we easily get that the infimum is obtained for  $\theta = \mathcal{S}\, \mathrm{arctanh}(\delta)$ and the bound is:
\begin{equation*}
    F(\delta,\lambda,\mc{S})= \frac{2d_\lambda}{\left(1+\delta\right)^{\frac{\mc S}{2}\left(1+\delta\right)}\left(1-\delta\right)^{\frac{\mc S}{2}\left(1-\delta\right)}}=\frac{2d_\lambda}{(1-\delta^2)^{\frac{S}{2}}}e^{-\delta\mc{S}\,\mathrm{arctanh}(\delta)}.
\end{equation*}


\end{proof}
\subsubsection{Symmetric  Haar random gate-sets}
\color{black}
In the following we develop new bounds which are based on the fact that we can calculate explicitly the expected values that appear in the master bound (see Section \ref{sec:inequalities_review}). In order to perform this calculation we need to know all moments $\delta_\lambda(n)$. To do this for $n \le d$ we use an explicit formula from Fact \ref{fact:fs_indicator} and for $n > d$ we use it simplified form from Lemma \ref{lemma:abuhamed_simplified}. We derive the bound in Theorem \ref{th:symmetric_master_bound} which, using Fact \ref{besel}, we next optimize to obtain the main result of this section, that is, Theorem \ref{th:simplified_master_bound}.
\color{black}


\begin{theorem}
\label{th:symmetric_master_bound}
Let $\mathcal{S}$ be a symmetric Haar random gate-set from $SU(d)$ \textcolor{black}{ and $\nu_\mc{S}$ a uniform measure}. Then

\begin{gather}\label{eq:master_bessels}
    \mathbb{P} \left( \delta(\nu_{\mc{S}},\lambda^s)\| \geq \delta \right) \leq d_{\lambda^s} \left[ \mathrm{inf}_{\theta>0}\; e^{-\theta \delta} F(\theta,\lambda^s, \mc{S}) + \mathrm{inf}_{\theta>0}\; e^{-\theta \delta} F(-\theta,\lambda^s, \mc{S}) \right],
\end{gather}
where
\begin{align}\label{eq:F}
    F(\theta,\lambda^s, \mc{S}) &= \left[ \frac{m_{\lambda^s}^0}{d_{\lambda^s}} e^{\frac{2\theta}{\mc{S}}} + \sum_{k=-d}^d \gamma_{\lambda^s}(k) I_{|k|}\left( \frac{2\theta}{\mc{S}} \right) \right]^\frac{\mc{S}}{2},\\
    \nonumber
    \gamma_{\lambda^s}(k) &= \begin{cases}
        1 - \frac{m^0_{\lambda^s}}{d_{\lambda^s}} &\text{if $k=0$},\\
        \frac{1}{d_{\lambda^s}} \sum_{\substack{\sigma \in S_d \\ \sigma \neq id}} \mathrm{sgn}(\sigma) m_{\lambda^s}\left( \frac{\rho - \sigma \cdot \rho}{k} \right) &\text{if $k\neq0$},
    \end{cases}
\end{align}
and $I_n(x)$ is $n$-th modified Bessel function of the first kind.
\end{theorem}
\begin{proof}
Clearly, the bound from Fact \ref{fact:master_bound} is of the form
\begin{equation*}
    \mathrm{inf}_{\theta>0}\; e^{-\theta \delta} \tilde{F}(\theta,\lambda^s, \mc{S}) + \mathrm{inf}_{\theta>0}\; e^{-\theta \delta} \tilde{F}(-\theta,\lambda^s, \mc{S}),
\end{equation*}
where
\begin{align*}
    \tilde{F}(\theta,\lambda^s, \mc{S}) &= \mathrm{tr} \exp \left( \sum_{k=1}^{\mc{S}/2} \log \textcolor{black}{\E{\mu}} e^{\frac{\theta}{\mc{S}} X_{U_k, \lambda^s}}\right), \\
    X_{U,\lambda^s} &= \pi_{\lambda^s}(U)+\pi_{\lambda^s}(U^{-1}).
\end{align*}
Thus our main objective is to compute the $\textcolor{black}{\E{\mu}} e^{\frac{\theta}{\mc{S}} X_{U, \lambda^s}}$. Using the binomial formula one easily gets:
\begin{gather}
    \label{eq:exp_val_power_of_X}
    \textcolor{black}{\E{\mu}} X_{U,\lambda^s}^n =
    \sum_{k=0}^{n} {n \choose k} \textcolor{black}{\E{\mu}} \pi_{\lambda^s}(U^{n-2k}) =
    \bbone_{\lambda^s} \left[\sum_{k=0}^{n} {n \choose k} \delta_{\lambda^s}(n-2k) \right].
\end{gather}
From Lemma \ref{lemma:abuhamed_simplified} we know that for $|n-2k| > d$ the value of $\delta_{\lambda^s}$ is $\frac{m_{\lambda^s}^0}{d_{\lambda^s}}$. Thus Eq. \eqref{eq:exp_val_power_of_X} becomes
\begin{gather}
    \nonumber
    \textcolor{black}{\E{\mu}} X_{U,\lambda^s}^n = \bbone_{\lambda^s} \left[\sum_{k=0}^{n} {n \choose k} \frac{m_{\lambda^s}^0}{d_{\lambda^s}} +  
    \sum_{k=\lceil \frac{n-d}{2} \rceil }^{\lfloor \frac{n+d}{2} \rfloor} {n \choose k} \left( \delta_{\lambda^s}(n-2k) - \frac{m_{\lambda^s}^0}{d_{\lambda^s}} \right)  \right] =\\
    \nonumber
    =\bbone_{\lambda^s} \left[ \frac{m_{\lambda^s}^0\, 2^n}{d_{\lambda^s}} + \sum_{k=\lceil \frac{n-d}{2} \rceil }^{\lfloor \frac{n+d}{2} \rfloor} {n \choose k} \gamma_{\lambda^s}(n-2k)  \right],
\end{gather}
where we assume ${n \choose k}=0$ for $n < k$ or $k < 0$.  We will consider separately cases for different parities of $d$ and $n$:\\

$\bullet \quad d = 2p$ and $n=2l$:
\begin{gather*}
    \textcolor{black}{\E{\mu}} X_{U,\lambda^s}^{2l} = \bbone_{\lambda^s} \left[ \frac{m_{\lambda^s}^0\, 2^{2l}}{d_{\lambda^s}} + 
    \sum_{k=l-p}^{l+p} {2l \choose k} \gamma_{\lambda^s}(2l-2k)  \right] =\\=
    \bbone_{\lambda^s} \left[ \frac{m_{\lambda^s}^0\, 2^{2l}}{d_{\lambda^s}} + 
    \sum_{k=-p}^{p} {2l \choose l+k} \gamma_{\lambda^s}(2k)  \right] ,
\end{gather*}

$\bullet \quad d = 2p$ and $n=2l+1$:
\begin{gather*}
    \textcolor{black}{\E{\mu}} X_{U,\lambda^s}^{2l+1} = \bbone_{\lambda^s} \left[ \frac{m_{\lambda^s}^0\, 2^{2l+1}}{d_{\lambda^s}} + 
    \sum_{k=l+1-p}^{l+p} {2l+1 \choose k} \gamma_{\lambda^s}(2l+1-2k)  \right] =\\=
    \bbone_{\lambda^s} \left[ \frac{m_{\lambda^s}^0\, 2^{2l+1}}{d_{\lambda^s}} + 
    \sum_{k=-p+1}^{p} {2l+1 \choose l+k} \gamma_{\lambda^s}(2k-1)  \right].
\end{gather*}
Then we have:
\begin{gather}
    \nonumber
    \textcolor{black}{\E{\mu}} e^{\frac{\theta}{\mc{S}}X_{U,\lambda^s}} = \bbone \left[ \frac{m^0_{\lambda^s}}{d_{\lambda^s}}e^{\frac{2\theta}{\mc{S}}} + \sum_{k=-p}^{p} \gamma_{\lambda^s}(2k) I_{|2k|}\left( \frac{2\theta}{\mc{S}} \right) + \sum_{k=-p+1}^{p} \gamma_{\lambda^s}(2k-1) I_{|2k-1|}\left( \frac{2\theta}{\mc{S}} \right) \right] =\\
    \label{eq:exp_even_d}
    =\bbone \left[ \frac{m^0_{\lambda^s}}{d_{\lambda^s}}e^{\frac{2\theta}{\mc{S}}} + \sum_{k=-d}^{d} \gamma_{\lambda^s}(k) I_{|k|}\left( \frac{2\theta}{\mc{S}} \right) \right].
\end{gather}

$\bullet \quad d = 2p+1$ and $n=2l$:
\begin{gather*}
    \textcolor{black}{\E{\mu}} X_{U,\lambda^s}^{2l} = \bbone_{\lambda^s} \left[ \frac{m_{\lambda^s}^0\, 2^{2l}}{d_{\lambda^s}} + 
    \sum_{k=l-p}^{l+p} {2l \choose k} \gamma_{\lambda^s}(2l-2k)  \right] =\\=
    \bbone_{\lambda^s} \left[ \frac{m_{\lambda^s}^0\, 2^{2l}}{d_{\lambda^s}} + 
    \sum_{k=-p}^{p} {2l \choose l+k} \gamma_{\lambda^s}(2k)  \right] ,
\end{gather*}

$\bullet \quad d = 2p+1$ and $n=2l+1$:
\begin{gather*}
    \textcolor{black}{\E{\mu}} X_{U,\lambda^s}^{2l+1} = \bbone_{\lambda^s} \left[ \frac{m_{\lambda^s}^0\, 2^{2l+1}}{d_{\lambda^s}} + 
    \sum_{k=l-p}^{l+p+1} {2l+1 \choose k} \gamma_{\lambda^s}(2l+1-2k)  \right] =\\=
    \bbone_{\lambda^s} \left[ \frac{m_{\lambda^s}^0\, 2^{2l+1}}{d_{\lambda^s}} + 
    \sum_{k=-p}^{p+1} {2l+1 \choose l+k} \gamma_{\lambda^s}(2k-1)  \right].
\end{gather*}
Then we have:
\begin{gather}
    \nonumber
    \textcolor{black}{\E{\mu}} e^{\frac{\theta}{\mc{S}}X_{U,\lambda^s}} = \bbone \left[ \frac{m^0_{\lambda^s}}{d_{\lambda^s}}e^{\frac{2\theta}{\mc{S}}} + \sum_{k=-p}^{p} \gamma_{\lambda^s}(2k) I_{|2k|}\left( \frac{2\theta}{\mc{S}} \right) + \sum_{k=-p}^{p+1} \gamma_{\lambda^s}(2k-1) I_{|2k-1|}\left( \frac{2\theta}{\mc{S}} \right) \right] =\\
    \label{eq:exp_odd_d}
    = \bbone \left[ \frac{m^0_{\lambda^s}}{d_{\lambda^s}}e^{\frac{2\theta}{\mc{S}}} + \sum_{k=-d}^{d} \gamma_{\lambda^s}(k) I_{|k|}\left( \frac{2\theta}{\mc{S}} \right) \right].
\end{gather}

Combining Eq. \ref{eq:exp_even_d} and Eq. \ref{eq:exp_odd_d} with Fact \ref{fact:master_bound} we get the desired result.

\end{proof}
Note that from Kostant formula \eqref{fact:kostant} the multiplicities $m^\mu_{\lambda^s}$ are polynomials in coefficients of $\lambda^s$ with degree up to $\frac{(d-2)(d-1)}{2}$ while from Weyl dimension formula \eqref{eq:weyl_dimension} the dimension $d_{\lambda^s}$ is a polynomial of degree $\frac{(d-1)d}{2}$. Thus for $\lambda^s$ with large coefficients the expression in bracket in Eq. \eqref{eq:master_bessels} is approximately equal to $\gamma_{\lambda^s}(0)I_0(\frac{2\theta}{\mc{S}})$. In order to proceed we need the following property of the modified Bessel functions.

\begin{fact}\label{besel}
\cite{bessel} For any $n\geq 0$ we have the following upper and lower bounds on the ratio of the modified Bessel functions
\begin{gather*}
\frac{x}{n-\frac{1}{2}+\sqrt{\left(n+\frac{1}{2}\right)^2+x^2}}    <\frac{I_n(x)}{I_{n-1}(x)}<\frac{x}{n-1+\sqrt{\left(n+1\right)^2+x^2}}.
\end{gather*}
\end{fact}
Combining Fact \ref{besel} with Theorem \ref{th:symmetric_master_bound} we arrive at
\begin{theorem}
\label{th:simplified_master_bound}
Let $\mathcal{S}$ be a symmetric Haar random gate-set from $SU(d)$ \textcolor{black}{ and $\nu_\mc{S}$ a uniform measure}. Then
\begin{gather*}
    \mathbb{P} \left( \delta(\nu_{\mc{S}},\lambda) \geq \delta\right) \leq d_{\lambda^s} \, e^{-\frac{\mc{S}\delta^2}{\sqrt{1-\delta^2}}} \left[ F\left(\frac{\mc{S}\delta}{\sqrt{1-\delta^2}},\lambda, \mc{S}\right) +  F\left(-\frac{\mc{S}\delta}{\sqrt{1-\delta^2}},\lambda, \mc{S}\right) \right],
\end{gather*}
 where $F(\cdot,\cdot,\cdot)$ is given by \eqref{eq:F}.
\end{theorem}
\begin{proof}
In order to find the best bound we need to determine $\theta$ that realizes
\begin{gather*}
    \mathrm{inf}_{\theta>0}\; e^{-\theta \delta} 
    \left[ \frac{m_{\lambda^s}^0}{d_{\lambda^s}} e^{\frac{2\theta}{\mc{S}}} + \sum_{k=-d}^d \gamma_{\lambda^s}(k) I_{|k|}\left( \frac{2\theta}{\mc{S}} \right) \right]^\frac{\mc{S}}{2}.
\end{gather*}
As finding minimum of the above functions is analytically intractable, in both cases we look for $\theta$ that minimizes $e^{-\theta\delta }I_0\left(\frac{2\theta}{\mc{S}}\right)^{\frac{\mc{S}}{2}}$. Taking the derivative with respect to $\theta$ we get
\begin{gather*}
\delta=\frac{I_1\left(x\right)}{I_0\left(x\right)},
\end{gather*}
where $x=\frac{2\theta}{\mc{S}}$. Using the upper bound for the ration of modified Bessel functions from Fact \ref{besel} we get 
\begin{gather*}
    \delta=\frac{x}{\sqrt{4+x^2}}\,\implies\, x=\frac{2\delta}{\sqrt{1-\delta^2}}.
\end{gather*}
The result follows.
\end{proof}

\color{black}
\subsection{Bernstein and master bounds for random circuits} 
\label{subsec:bounds-for-circuits}
Let us consider a scenario where we draw $m$ independent gates from the exact $t_0$-design $\{ \mathcal{D}, \nu_\mathcal{D} \}$ and combine them into a circuit $\mathbf{U} = (U_1, ..., U_m)$. We repeat this procedure some number of times and in this way we construct a set of random circuits $\mathcal{R} \subset \mathcal{D}^{\times m}$. Then for $f \in \mathcal{H}_{t, m}$ we ask how much averaging of $f$ over $\mathcal{R}$ differs from averaging over all circuits $\mathcal{D}^{\times m}$. More precisely we want to bound the tail probability
\begin{equation*}
    \mathbb{P} ( \| T_{\nu_\mathcal{D}^{\times m}, t} - T_{\nu_\mathcal{R}, t} \| \ge \delta),
\end{equation*}
for $\nu_\mathcal{R}$ uniform, $0< \delta < 1$ and $t \le t_0$. Note that since $\mathcal{D}$ is an exact $t_0$-design for any $t \le t_0$ we have:
\begin{equation*}
    T_{\nu_\mathcal{D}^{\times m}, t} = \Ec{\nu_\mathcal{D}}{1} ... \Ec{\nu_\mathcal{D}}{m} \bigotimes_{i=1}^m U_i^{t, t} = \bigotimes_{i=1}^m \left( \Ec{\nu_\mathcal{D}}{i} U_i^{t, t} \right) = \bigotimes_{i=1}^m T_{\mu, t} = T_{\mu, t}^{\otimes m}.
\end{equation*}
Thus
\begin{equation*}
    \mathbb{P} ( \| T_{\nu_\mathcal{D}^{\times m}, t} - T_{\nu_\mathcal{C}, t} \| \ge \delta) = \mathbb{P} ( \delta(\nu_\mathcal{C}, t) \ge \delta) \le \sum_{\bm{\lambda} \in \Lambda_{t, m}} \mathbb{P} ( \delta(\nu_\mathcal{C}, \bm{\lambda} ) \ge \delta ).
\end{equation*}
With this we can prove Bernstein and master bounds for random circuits in a very similar way to the case with random gates. For example, consider calculation of moments of $\pi_{\bm{\lambda}}(\mathbf{U})$ for $n \cdot t \le t_0$:
\begin{equation*}
    \Ec{\nu_\mathcal{D}}{1} ... \Ec{\nu_\mathcal{D}}{m} \pi_{\bm{\lambda}}(\mathbf{U})^n = \bigotimes_{i=1}^m \left( \Ec{\nu_D}{i} \pi_{\lambda_i}(U_i)^n \right) = \prod_{i=1}^m \delta_{\lambda_i}(n) \bbone.
\end{equation*}
Therefore Bernstein, symmetric Bernstein and master bounds for random circuits are almost exactly the same as for random gates with only small changes summarized in the below diagram:
\begin{align*}
    \mathcal{S} &\longrightarrow \mathcal{C},\\
    \sum_{\lambda \in \Lambda_t} &\longrightarrow \sum_{\bm{\lambda} \in \Lambda_{t, m}},\\
\end{align*}
and then for $\bm{\lambda} = (\lambda_1, ..., \lambda_m) \in \Lambda_{t, m}$:
\begin{align*}
    \delta_\lambda(n) &\longrightarrow \delta_{\bm{\lambda}} (n) :=  \prod_{i=1}^m \delta_{\lambda_i}(n),\\
    d_\lambda &\longrightarrow d_{\bm{\lambda}} := \prod_{i=1}^m d_{\lambda_i}.
\end{align*}
In case of symmetric master bound we have to additionally substitute
\begin{equation*}
    m_{\lambda^s}^0 \longrightarrow m^0_{\bm{\lambda}^s} := \prod_{i=1}^m m_{\lambda_i^s}^0,
\end{equation*}
and redefine $\gamma_{\lambda^s}(n)$ to
\begin{equation*}
    \gamma_{\bm{\lambda}^s}(k) = \begin{cases}
        1 - \prod_{i=1}^m \frac{m^0_{\lambda^s_i}}{d_{\lambda^s_i}} &\text{if $k=0$},\\
        \prod_{i=1}^m \left[ \frac{1}{d_{\lambda^s_i}} \sum_{\sigma \in S_d} \mathrm{sgn}(\sigma) m_{\lambda^s_i}\left( \frac{\rho - \sigma \cdot \rho}{k} \right) \right]- \prod_{i=1}^m \frac{m^0_{\lambda^s_i}}{d_{\lambda^s_i}} &\text{if $k\neq0$}.
    \end{cases}
\end{equation*}

\color{black}

\subsection{Concentration of $\delta(\nu_\mc{S},t)$ and $\delta(\nu_\mc{S},\lambda)$ around their expected values}
In this section we {\color{black} derive Theorems \ref{th:intro_master_bound-main3} and \ref{th:intro_beam} from Introduction about the concentration} properties of $\delta(\nu_\S, t)$ and $\delta(\nu_\S, \lambda)$ using Fact \ref{fact:haar_inequality}. 
\begin{theorem}\label{theorem:Haar}
Let $\mathcal{S}\subset SU(d)$ be a Haar random gate-set. Then
\begin{gather*}
\mathbb{P}\left(\delta(\nu_{\mathcal{S}}, t) \geq \textcolor{black}{\E{\mu}} \delta(\nu_{\mathcal{S}},t)+\alpha\right)\leq \exp\left({\frac{-d\mathcal{S}\alpha^2}{32t^2}}\right).
\end{gather*}
\end{theorem}
\begin{proof}
Let $\mc{S}\subset SU(d)$ be any set of cardinality $n>1$. For the convenience we denote by $\mathcal{S}_1=\{U_1,\ldots,U_n\}$, and  $\mathcal{S}_2=\{V_1,\ldots,V_n\}$ two exemplary sets $\mc{S}$. Let 
\begin{gather*}
    F(U_1,\ldots,U_n)=\|T_{\nu_{\mc{S}_1},t}-T_{\mu,t}\|.
\end{gather*}
We need to show that there is a constant $L$ such that for any $\mc{S}_1$ and $\mc{S}_2$ we have 
\begin{gather*}
    | F(U_1,\ldots,U_n)-F(V_1,\ldots,V_n)|\leq L \left(\sum_{i=1}^n\|U_i-V_i\|^2_F\right)^{\frac{1}{2}}.
\end{gather*}
In order to determine the value of $L$ we go through the following chain of inequalities
\begin{gather*}
    \left|\|T_{\nu_{\mc{S}_1},t}-T_{\mu,t}\|-\|T_{\nu_{\mc{S}_2},t}-T_{\mu,t}\|\right|\leq\|T_{\nu_{\mc{S}_1},t}-T_{\nu_{\mc{S}_2},t}\|\leq \frac{1}{n}\sum_{i=1}^{n} \|U_i^{t,t}-V_i^{t,t}\|\leq\nonumber\\\leq \frac{2t}{n}\sum_{i=1}^{n} \|U_i-V_i\|\leq\frac{2t}{n}\sum_{i=1}^{n} \|U_i-V_i\|_{F}\leq \frac{2t}{\sqrt{n}}\left(\sum_{i=1}^n\|U_i-V_i\|^2_F\right)^{\frac{1}{2}}.
\end{gather*}
Thus the Lipschitz constant is $L=\frac{2t}{\sqrt{\mc{S}}}$. Knowing the value of $L$ we use Fact \ref{fact:haar_inequality} and obtain the result.
\end{proof}

\begin{theorem}\label{theorem:Haar_lambda}
Let $\mathcal{S}\subset SU(d)$ be a Haar random gate-set. Then
\begin{gather*}
\mathbb{P}\left(\delta(\nu_{\mathcal{S}},\lambda)\geq \textcolor{black}{\E{\mu}}\delta(\nu_{\mathcal{S}},\lambda)+\alpha\right)\leq \exp\left({\frac{-d\mathcal{S}\alpha^2}{2\pi^2\|\lambda\|_1^2}}\right).
\end{gather*}
\end{theorem}
\begin{proof}
Let $\mc{S}\subset SU(d)$ be any set of cardinality $n>1$. For the convenience we denote by $\mathcal{S}_1=\{U_1,\ldots,U_n\}$, and  $\mathcal{S}_2=\{V_1,\ldots,V_n\}$ two exemplary sets $\mc{S}$. Let 
\begin{gather*}
    F(U_1,\ldots,U_n)=\|T_{\nu_{\mc{S}_1},\lambda}\|.
\end{gather*}
We need to show that there is a constant $L$ such that for any $\mc{S}_1$ and $\mc{S}_2$ we have 
\begin{gather*}
    | F(U_1,\ldots,U_n)-F(V_1,\ldots,V_n)|\leq L \left(\sum_{i=1}^n\|U_i-V_i\|^2_F\right)^{\frac{1}{2}}.
\end{gather*}
In order to determine the value of $L$ we go through the following chain of inequalities
\begin{gather*}
    \left|\|T_{\nu_{\mc{S}_1},\lambda}\|-\|T_{\nu_{\mc{S}_2},\lambda}\|\right|\leq \frac{1}{n}\sum_{i=1}^{n} \|\pi_{\lambda}({U_i})-\pi_{\lambda}({V_i})\|\leq\nonumber\\\leq  \frac{\pi\|\lambda\|_1}{2n}\sum_{i=1}^{n} \|U_i-V_i\|\leq\frac{\pi\|\lambda\|_1}{2\sqrt{n}}\left(\sum_{i=1}^n\|U_i-V_i\|^2_F\right)^{\frac{1}{2}},
\end{gather*}
where in the second inequality we used Lemma \ref{lemma:norm_bound}. Thus the Lipschitz constant is $L=\frac{\pi\|\lambda\|_1}{2\sqrt{\mc{S}}}$. Knowing the value of $L$ we use Fact \ref{fact:haar_inequality} and obtain the result.
\end{proof}



\subsubsection{$d$-mode beamsplitters built from random $2$-mode beamsplitters}\label{sec:2-mode-d-mode}

In this section we consider the Hilbert space $\mathcal{H}=\mc{H}_1\oplus\ldots\oplus\mc{H}_d$, where $\mc{H}_k\simeq\mathbb{C}$, $d>2$. We will call spaces $\mc{H}_k$ modes. For a matrix $B\in SU(2)$, which we call a $2$-mode beamsplitter, we define matrices $B^{ij}$, $i\neq j$, to be the matrices that act on a $2$-dimensional subspace $\mc{H}_i\oplus\mc{H}_j\subset \mc{H}$ as $B$ and on the other components of $\mc{H}$ as the identity. This way a matrix $B\in SU(2)$ gives $d(d-1)$ matrices in $SU(d)$. Applying this procedure to a Haar random gate-set set $S\subset SU(2)$ we obtain random gate-set $S^d$ (see \cite{Reck,sawicki1}). We are interested in efficiency of $\mc{S}^d$. 

\begin{theorem}\label{theorem:beam}
Let $\mc{S}\subset SU(2)$ be a Haar random gate-set and $\mc{S}^d\subset SU(d)$ the corresponding $d$-mode gate-set. Then 
\begin{gather*}
\mathbb{P}\left(\delta(\nu_{\mathcal{S}^d},t)\geq \textcolor{black}{\E{\mu}}\delta(\nu_{\mathcal{S}^d},t)+\alpha\right)\leq \exp\left({\frac{-\mathcal{S}\alpha^2}{16t^2}}\right).
\end{gather*}
\end{theorem}
\begin{proof}
Let $\mc{S}_1,\mc{S}_2\subset SU(2)$ be two gate-sets of the same size $n$, i.e. $\mathcal{S}_1=\{U_1,\ldots,U_n\}$, and  $\mathcal{S}_2=\{V_1,\ldots,V_n\}$. We denote by $\mathcal{S}_1^d$, and  $\mathcal{S}_2^d$ the corresponding $d$-mode gate-sets
\begin{gather*}
    \mathcal{S}_1^d=\{U_k^{i,j}|k\in \{1,\ldots,n\},\,i,j\in \{1,\ldots,d\}\,,i\,\neq j\},\\
    \mathcal{S}_2^d=\{V_k^{i,j}|k\in \{1,\ldots,n\},\,i,j\in \{1,\ldots,d\}\,,\,i\neq j\}.
\end{gather*}
Similarly as in the proofs of Theorems \ref{theorem:Haar} and \ref{theorem:Haar_lambda} we calculate,
\begin{gather*}
    \left|\|T_{\nu_{\mc{S}_1^d},t}-T_{\mu,t}\|-\|T_{\nu_{\mc{S}_2^d},t}-T_{\mu,t}\|\right|
    \leq\|T_{\nu_{\mc{S}_1^d},t}-T_{\nu_{\mc{S}_2^d},t}\|\leq\\\nonumber\leq \frac{1}{d(d-1)}\sum_{1\leq i\neq j\leq d}\frac{1}{n}\sum_{k=1}^{n} \|(U_k^{i,j})^{t,t}-(V_k^{i,j})^{t,t}\|\leq\frac{1}{d(d-1)}\sum_{1\leq i\neq j\leq d}\frac{2t}{n}\sum_{k=1}^{n} \|U_k-V_k\|_{F}\leq\\\nonumber\frac{1}{d(d-1)}\sum_{1\leq i\neq j\leq d} \frac{2t}{\sqrt{n}}\left(\sum_{k=1}^n\|U_k-V_k\|^2_F\right)^{\frac{1}{2}}=\frac{2t}{\sqrt{n}}\left(\sum_{k=1}^n\|U_k-V_k\|^2_F\right)^{\frac{1}{2}}.
\end{gather*}
Thus the Lipschitz constant is $L=\frac{2t}{\sqrt{\mc{S}}}$. Knowing the value of $L$ we use Fact \ref{fact:haar_inequality} and obtain the result.
\end{proof}

As a conclusion we see that a Haar random gate-set $\mc{S}\subset SU(2)$ gives the gate-set $S^d\subset SU(d)$ for which $\delta(\nu_{\mathcal{S}^d},t)$ has the same concentration rate around the mean as a Haar random gate-set $\mc{S}^\prime\subset SU(d)$ of size:
\begin{gather*}
    \mc{S}^\prime=\frac{2\mc{S}}{d}.
\end{gather*}
We note, however, that using this approach one cannot say anything about the relationship between $\textcolor{black}{\E{\mu}}\delta(\nu_{\mathcal{S}^d},t)$ and $\textcolor{black}{\E{\mu}}\delta(\nu_{\mathcal{S}^\prime},t)$.

\section{Comparison of bounds}\label{sec:inequalities_comparison}

\begin{figure}[htp]
    \centering
    \includegraphics[scale=0.9]{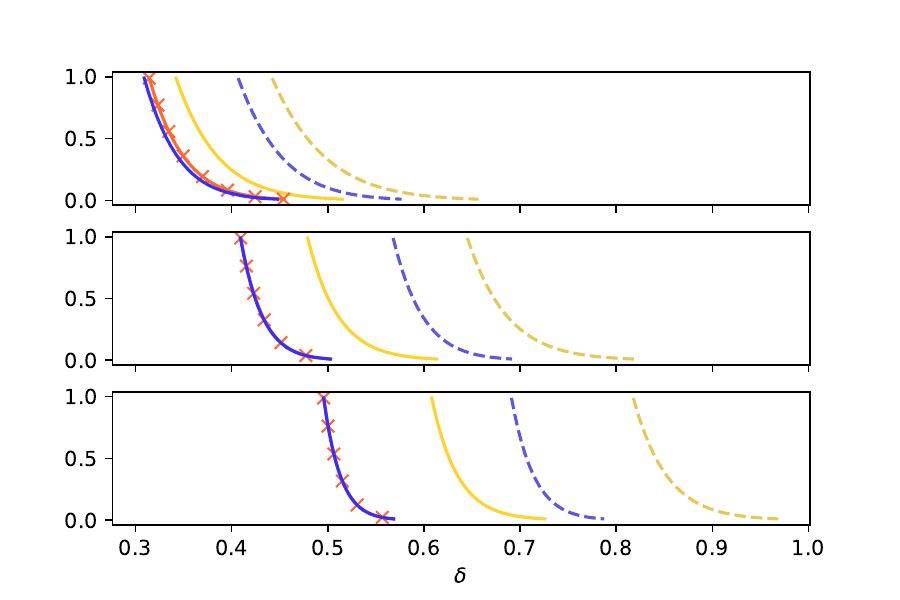}
    \caption{Upper bounds on $\mathbb{P}(\| T_{\nu_\mc{S}, t} \| \ge \delta)$ for $d=2$, $\mc{S}=50$ or in symmetric case $\mc{S} = 2 \times 50$ and different $t$: top - $5$, middle - $50$, bottom - $500$. Master - dashed blue,  symmetric master - solid blue, simplified symmetric master - solid orange with x-markers, Bernstein - dashed yellow and symmetric Bernstein - solid yellow.}
    \label{fig:t5_50_500}
\end{figure}

\begin{figure}
    \centering
    \includegraphics[scale=0.9]{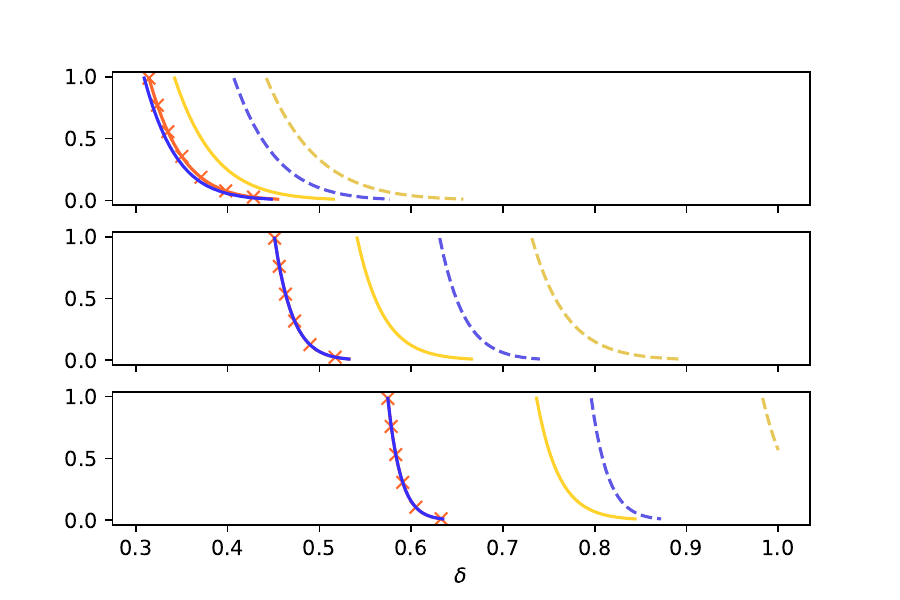}
    \caption{Upper bounds on $\mathbb{P}(\| T_{\nu_\mc{S}, t} \| \ge \delta)$ for $t=5$, $\mc{S}=50$ or in symmetric case $\mc{S} = 2 \times 50$ and different $d$: top - $2$, middle - $4$, bottom - $8$. Master - dashed blue,  symmetric master - solid blue, simplified symmetric master - solid orange with x-markers, Bernstein - dashed yellow and symmetric Bernstein - solid yellow.}
    \label{fig:d2_4_8}
\end{figure}

\begin{figure}
    \centering
    \includegraphics[scale=0.9]{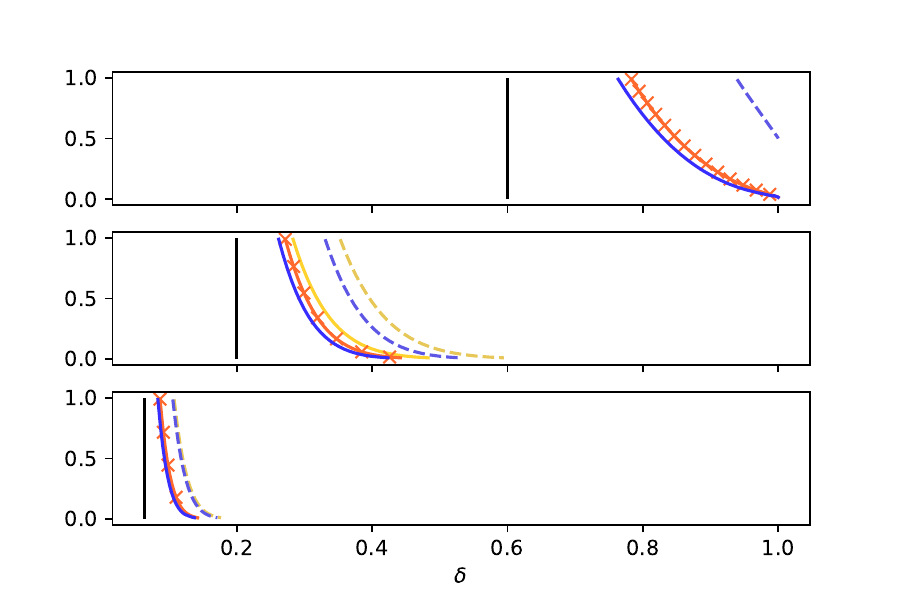}
    \caption{Upper bounds on $\mathbb{P}(\| T_{\nu_\mc{S}, t} \| \ge \delta)$ for $d=2$, $t=2$ and different $\mc{S}$: top - $5$, middle - $50$, bottom - $500$ or in symmetric case $\mc{S}$: top - $2 \times 5$, middle - $2 \times 50$, bottom - $2 \times 500$. Master - dashed blue,  symmetric master - solid blue, simplified symmetric master - solid orange with x-markers, Bernstein - dashed yellow and symmetric Bernstein - solid yellow. Black vertical lines indicate the value of $\delta_\mathrm{opt}(\mc{S})$.}
    \label{fig:S_10_100_1000}
\end{figure}

\begin{figure}
    \centering
    \includegraphics[scale=0.9]{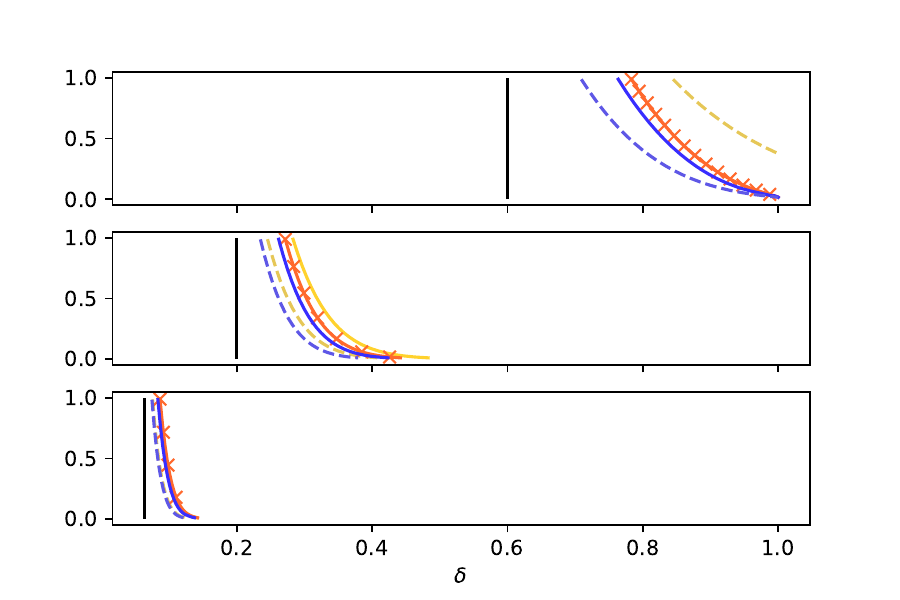}
    \caption{\color{black} Upper bounds on $\mathbb{P}(\| T_{\nu_\mc{S}, t} \| \ge \delta)$ for $d=2$, $t=2$ and different $\mc{S}$: top - $10$, middle - $100$, bottom - $1000$ or in symmetric case $\mc{S}$: top - $2 \times 5$, middle - $2 \times 50$, bottom - $2 \times 500$. Master - dashed blue,  symmetric master - solid blue, simplified symmetric master - solid orange with x-markers, Bernstein - dashed yellow and symmetric Bernstein - solid yellow. Black vertical lines indicate the value of $\delta_\mathrm{opt}(\mc{S})$.}
    \label{fig:S_10_100_1000_same_sizes}
\end{figure}

In this section we use numerical results to compare derived bounds for various values of $t$ (Fig. \ref{fig:t5_50_500}), $d$ (Fig. \ref{fig:d2_4_8}) and $\mc{S}$ (Fig. \ref{fig:S_10_100_1000} \textcolor{black}{and Fig. \ref{fig:S_10_100_1000_same_sizes}}). Throughout this section we will use the following convention:
\begin{itemize}
    \item Bounds using Theorem \ref{th:bernstein_bound} will be called  Bernstein bounds and symmetric Bernstein bounds. They will be plotted with a dashed yellow line and a solid yellow line respectively.
    \item Bounds using Theorem \ref{th:master_bound} will be called master bounds and they will be plotted with a dashed blue line.
    \item Bounds using Theorem \ref{th:symmetric_master_bound} will be called symmetric master bounds and they will be plotted with a solid blue line.
    \item Bounds using Theorem \ref{th:simplified_master_bound} will be called simplified symmetric master bound and they will be plotted with solid orange line with crosses.
\end{itemize}
\color{black} We assume that gate-sets are $(c \cdot t)$-random with $c$ chosen in such a way that the Theorems mentioned above can be applied to $\delta(\nu_\mathcal{S}, t)$, that is:
\begin{equation*}
    c = \begin{cases}
        1 & \text{for Bernstein and master bounds,}\\
        2 & \text{for symmetric Bernstein bounds,} \\
        \infty & \text{for symmetric master bounds.}
    \end{cases}
\end{equation*}
\color{black}

\begin{figure}
    \centering
    \includegraphics[scale=0.75]{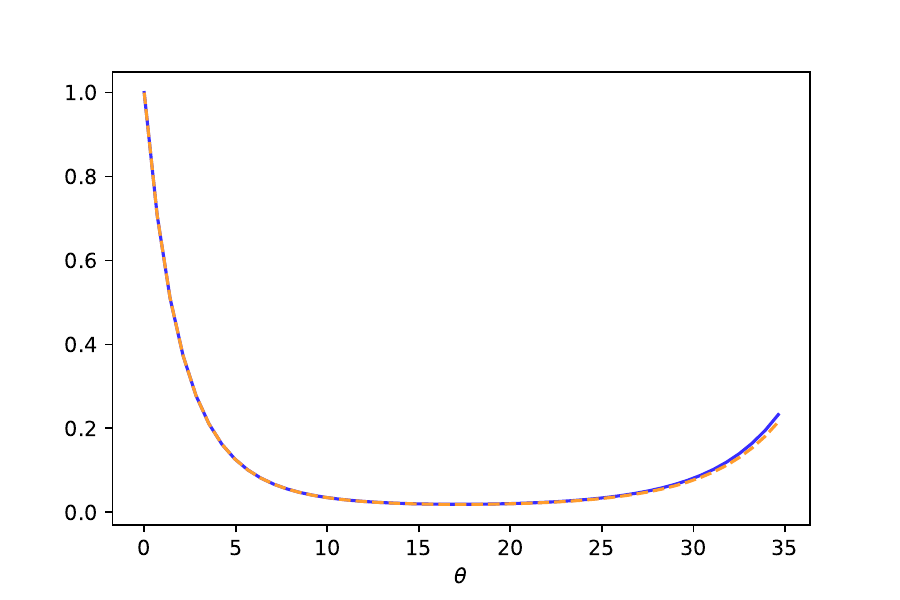}
    \caption{Functions under the infimum from Eq. \eqref{eq:master_bessels} for $\mc{S}=30$ and $\delta=0.5$.}
    \label{fig:master_bound_functions}
\end{figure}

First conclusion from Figures \ref{fig:t5_50_500}, \ref{fig:d2_4_8}, \ref{fig:S_10_100_1000} \textcolor{black}{and \ref{fig:S_10_100_1000_same_sizes}} is that the \textcolor{black}{(symmetric)} master bound is \textcolor{black}{tighter than the (symmetric) Bernstein bound. The difference} gets more pronounced with bigger $t$ and $d$ and smaller $\mc{S}$. Next, note that simplified symmetric master bound is almost identical with symmetric master bound what implies that our guess in derivation of Theorem \ref{th:simplified_master_bound} was very close to optimal, even for small $t$. This can be explained by the fact that the functions under the infimum from \eqref{eq:master_bessels} have the derivatives very close to zero in a quite wide interval near the minimum (see Fig. \ref{fig:master_bound_functions}). Thus the range of close to optimal guesses for infimum is quite wide as well.

We also analyze what is (according to our bounds) the gain of adding inverses to random gate-sets i.e. by making them symmetric. Figures \ref{fig:t5_50_500}, \ref{fig:d2_4_8} and \ref{fig:S_10_100_1000} indicate that bounds for set with inverses are tighter but the difference decreases with growing $\mc{S}$. \color{black} On the other hand, when we compare gate-sets of the same size (see Fig. \ref{fig:S_10_100_1000_same_sizes}) then bounds for sets without inverses are better. Thus, we conclude that new random gates improve $\delta(\nu_\mathcal{S}, t)$ more significantly than additional inverses of gates that were already in the set.

\color{black}

Next we check what are minimal sizes of $\mc{S}$ required to obtain $\delta$-approximate $t$-design, with probability at least $P$ according to the master bound. For \textcolor{black}{$t-$}random gate-sets the formula can be easily obtained from Theorem \ref{th:master_bound-main1} and reads:

\begin{gather}\label{eq:S_bound}
    \mc{S}\geq \frac{2\left(\log\left(2\sum_{\lambda \in \Lambda_t}d_\lambda\right)-\log\left(1-P\right)\right)}{\log\left(\left(1 + \delta\right)^{1+\delta}\left(1-\delta\right)^{1-\delta}\right)}.
\end{gather}

For symmetric Haar random gate-sets, however, it is much more difficult to obtain analogous formula using Theorem \ref{th:master_bound-main2}. Nevertheless the numerical calculations of Table \ref{tab:S_on_t} suggest that the required number of gates is around half of \eqref{eq:S_bound} plus inverses. 

\begin{figure}[htp]
    \centering
    \includegraphics[scale=0.8]{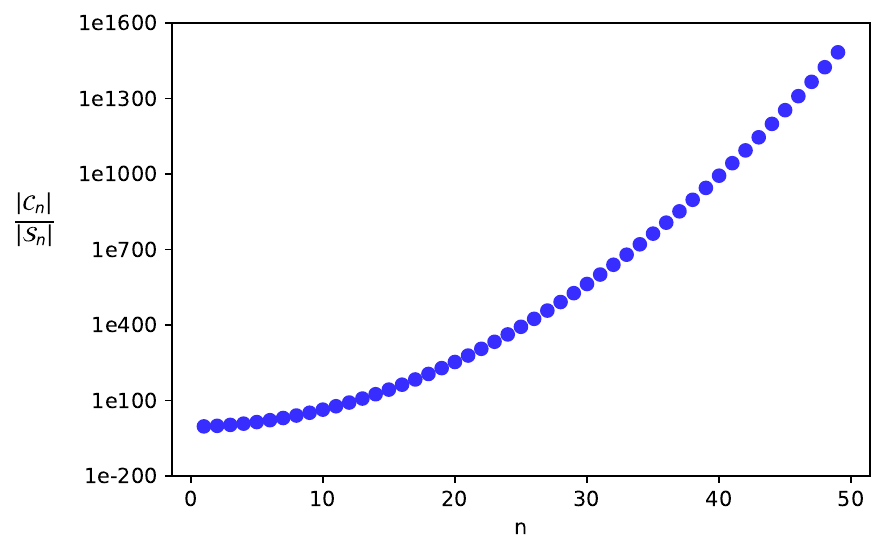}
    \caption{\textcolor{black}{Ratio of the cardinality of the set of $n$-qubit Clifford gates $\mathcal{C}_n$ and the minimal cardinality of the $t$-random gate-set $\mathcal{S}_n \subset U(2^n)$ that forms a $0.01$-approximate $2$-design with the probability at least $0.99$.}}
    \label{fig:comparison}
\end{figure}

For $d=2^n$ and  $t=2$, $P=0.99$, $\delta=0.01$, it is interesting to compare the number of gates given by \eqref{eq:S_bound}  with the number of elements of the $n$-qubit Clifford group, $\mathcal{C}_n$, which is \textcolor{black}{known} to be an exact $2$-design. Using Lemma \ref{lemma:partitions} one easily checks that for $t=2$ and $d\geq 4$ the set $\Lambda_{2}$ has exactly $p(2)^2+p(1)^2=5$ distinct $\lambda$'s. They are given by
\begin{gather*}
    \lambda_1=(1,0,\ldots,0,-1),\,\| \lambda_1\|_1=2,\,d_{\lambda_1}=d^2-1,\\\nonumber
    \lambda_2=(2,0,\ldots,0,-2),\,\| \lambda_2\|_1=4,\,d_{\lambda_2}=\frac{d^2(d-1)(d+3)}{4},\\\nonumber
    \lambda_3=(2,0,\ldots,0,-1,-1),\,\| \lambda_3\|_1=4,\,d_{\lambda_3}=\frac{(d^2-1)(d^2-4)}{4},\\\nonumber
    \lambda_4=(1,1,0,\ldots,0,-1,-1),\,\| \lambda_4\|_1=4,\,d_{\lambda_4}=\frac{d^2(d-3)(d+1)}{4},\\\nonumber
    \lambda_5=(1,1,0,\ldots,0,-2),\,\| \lambda_5\|_1=4,\,d_{\lambda_5}=\frac{(d^2-1)(d^2-4)}{4}.\nonumber
\end{gather*}
Thus $\sum_{\lambda \in \Lambda_2}d_\lambda=d^4-3d^2+1$. Using these results we find that  {\color{black}$t$-}random gate-set $\mc{S}_n\subset SU(2^n)$, $n\geq 2$, with
\begin{gather}
 \label{eq:S_bound1}
    \mc{S}_n\geq \left\lceil 2\cdot10^4\left(\log\left(2^{4n+1}-3\cdot2^{2n+1}+2\right)+4.61\right)\right\rceil.
\end{gather}
forms $0.01$-approximate $2$-design with the probability $P=0.99$.
On the other hand the cardinality of $\mathcal{C}_n$ is  given by \cite{NC00}
\begin{gather*}
    |\mathcal{C}_n|=2^{n^2+2n}\prod_{j=1}^n\left(4^j-1\right).
\end{gather*}
Figure \ref{fig:comparison} shows $\frac{|\mc{C}_n|}{\mc{S}_n}$ for up to $50$ qubits. The ratio $\frac{|\mc{C}_n|}{\mc{S}_n}$ grows at least exponentially with $n$. 
In fact, one can easily see, that $\sum_{\lambda \in \Lambda_t}d_\lambda\leq d^{2t}$. Thus \eqref{eq:S_bound} grows in a logarithmic way with $d$ and linearly with $t$ and {\color{black}$t$-}random gate-set $\S$ with

\begin{gather*}
\mc{S}\geq \frac{2\left(2t\log\left(d\right)+\log(2)-\log\left(1-P\right)\right)}{\log\left(\left(1+\delta\right)^{1+\delta}\left(1-\delta\right)^{1-\delta}\right)},
\end{gather*}
forms $\delta$-approximate $t$-design with the probability $P$. Moreover, for a small $\delta$ we can further reduce this formula using $\log\left(\left(1+\delta\right)^{1+\delta}\left(1-\delta\right)^{1-\delta}\right)\simeq \delta^2$. Thus we get that $S$ scales like $O( \delta^{-2}(t\log(d)-\log(1-P)))$.

We also note that if $\nu_{\mathcal{S}}$ is a $\delta$-approximate $t$-design with probability bounded by $1-\epsilon$. Then $\nu_{\mathcal{S}}^{\ast l}$, whose support are all circuits of depth $l$ built from gates form $\mathcal{S}$, is a $\delta^l$-approximate $t$-design with probability bounded by $1-\epsilon$. Thus one can construct first $1/2$-approximate $t$-design  and then by building all possible circuits of length $l$ change it to $1/2^l$-approximate $t$-design. Table \ref{tab:S_on_t}) presents numerical calculations for various $t$ and $d$.

\begin{table}[htp]
    \begin{center}
    \rotatebox{-90}{
    \begin{tabular}{|c|c|c|c|c|c|c|c|c|c|}
        \hline
         \multicolumn{3}{|c|}{t} & $2$ & $3$ & $4$ & $5$ & $20$ & $500$ & $5000$\\\hline
         \multirow{4}{*}{$d=2$} 
            & \multirow{2}{*}{Master} 
                & $\mathcal{S}$ & $57$ & $62$ & $65$ & $68$ & $88$ & $136$ & $171$ \\\cline{3-10}
                && $\mathcal{S}$-symmetric & $2 \times 36$ & $2 \times 37$ & $2 \times 39$ & $2 \times 40$ & $2 \times 46$ & $2 \times 67$ & $2 \times 84$ \\\cline{2-10}
            & \multirow{2}{*}{Bernstein}
                & $\mathcal{S}$ & $69$ & $75$ & $80$ & $83$ & $107$ & $166$ & $209$\\\cline{3-10}
                && $\mathcal{S}$-symmetric & $2 \times 47$ & $2 \times 50$ & $2 \times 52$ & $2 \times 53$ & $2 \times 64$ & $2 \times 95$ & $2 \times 120$\\\hline
        \multirow{4}{*}{$d=4$}
            & \multirow{2}{*}{Master}
                & $\mathcal{S}$ & $82$ & $96$ & $108$ & $118$ & $195$\\\cline{3-8}
                && $\mathcal{S}$-symmetric & $2 \times 41$ & $2 \times 47$ & $2 \times 53$ & $2 \times 58$ & $2 \times 95$\\\cline{2-8}
            & \multirow{2}{*}{Bernstein}
                & $\mathcal{S}$ & $100$ & $117$ & $132$ & $144$ & $238$ \\\cline{3-8}
                && $\mathcal{S}$-symmetric & $2 \times 58$ & $2 \times 67$ & $2 \times 76$ & $2 \times 82$ & $2 \times 136$ \\\cline{1-8}
        \multirow{4}{*}{$d=8$} 
            & \multirow{2}{*}{Master}
                & $\mathcal{S}$ & $104$ & $129$ & $154$ & $175$ \\\cline{3-7}
                && $\mathcal{S}$-symmetric & $2 \times 51$ & $2 \times 63$ & $2 \times 75$ & $2 \times 85$\\\cline{2-7}
            & \multirow{2}{*}{Bernstein}
                & $\mathcal{S}$ & $127$ & $158$ & $187$ & $213$ \\\cline{3-7}
                && $\mathcal{S}$-symmetric & $2 \times 73$ & $2 \times 90$ & $2 \times 107$ & $2 \times 122$ \\\cline{1-7}
        \multirow{3}{*}{$d=16$} 
            & \multirow{1}{*}{Master}
                & $\mathcal{S}$ & $126$ & $162$ & $197$ & $229$\\\cline{2-7}
            & \multirow{2}{*}{Bernstein}
                & $\mathcal{S}$ & $153$ & $197$ & $240$ & $280$ \\\cline{3-7}
                && $\mathcal{S}$-symmetric & $2 \times 88$ & $2 \times 113$ & $2 \times 137$ & $2 \times 160$ \\\cline{1-7}
        \multirow{3}{*}{$d=32$} 
            & \multirow{1}{*}{Master}
                & $\mathcal{S}$ & $147$ & $194$ & $239$ & $282$ \\\cline{2-7}
            & \multirow{2}{*}{Bernstein}
                & $\mathcal{S}$ & $179$ & $236$ & $292$ & $345$ \\\cline{3-7}
                && $\mathcal{S}$-symmetric & $2 \times 103$ & $2 \times 135$ & $2 \times 167$ & $2 \times 197$ \\\cline{1-7}
        \multirow{3}{*}{$d=64$}
            & \multirow{1}{*}{Master}
                & $\mathcal{S}$ & $168$ & $226$ & $282$ & $336$\\\cline{2-7}
            & \multirow{2}{*}{Bernstein}
                & $\mathcal{S}$ & $205$ & $275$ & $344$ & $410$ \\\cline{3-7}
                && $\mathcal{S}$-symmetric & $2 \times 117$ & $2 \times 158$ & $2 \times 197$ & $2 \times 234$ \\\cline{1-7}
        
    \end{tabular}
    }
    \end{center}
    \caption{Minimal sizes of Haar random gate-sets $\mathcal{S}$ required to obtain $\frac{1}{2}$-approximate $t$-design with probability at least $0.99$.}
    \label{tab:S_on_t}
\end{table}


\section*{Acknowledgments}
This research was funded by the National Science Centre, Poland under the grant OPUS: UMO-2020/37/B/ST2/02478 and supported in part by PLGrid Infrastructure.

\raggedright
\printbibliography

\end{document}